# High Frequency Lead/lag Relationships
# Empirical facts

Nicolas Huth*†     Frédéric Abergel †

January 10, 2012

**Abstract**

Lead/lag relationships are an important stylized fact at high frequency. Some assets follow the path of others with a small time lag. We provide indicators to measure this phenomenon using tick-by-tick data. Strongly asymmetric cross-correlation functions are empirically observed, especially in the future/stock case. We confirm the intuition that the most liquid assets (short intertrade duration, narrow bid/ask spread, small volatility, high turnover) tend to lead smaller stocks. However, the most correlated stocks are those with similar levels of liquidity. This lead/lag phenomenon is not constant throughout the day, it shows an intraday seasonality with changes of behaviour at very specific times such as the announcement of macroeconomic figures and the US market opening. These lead/lag relationships become more and more pronounced as we zoom on significant events. We reach 60% of accuracy when forecasting the next midquote variation of the lagger using only the past information of the leader, which is significantly better than using the information of the lagger only. However, a naive strategy based on market orders cannot make any profit of this effect because of the bid/ask spread.

## Introduction

The standard financial theory assumes that there is no arbitrage on financial markets[1][9]. In particular, it does not allow for predictability of asset returns. As a result, lead/lag relationships (assets driving others in advance) should not exist according to this theory. Figure 1 plots the cross-correlation function between the daily returns of the French equity index CAC40[2] (.FCHI) and those of the French stock Renault (RENA.PA), which is part of the CAC40, between 2003/01/02 and 2011/03/04[3]. The cross-correlation function is indeed close to a Dirac delta function (times the daily correlation). On a daily time scale, the absence of lead/lag relationships thus seems to be quite reasonable.

The availability of high frequency financial data allows us to zoom on microscopic fluctuations of the order flow. In this paper, we study the existence of lead/lag relationships between assets at fine time scales. Lead/lag relationships are measured with the Hayashi-Yoshida cross-correlation estimator [12, 13]. This estimator deals with the issue of asynchronous trading and makes use of all the available tick-by-tick data, so that we can theoretically measure lags down to the finest time scale. We report evidence of highly asymmetric cross-correlation functions as a witness of lead/lag relationships. These are not statistical artefacts due to differences in levels of trading activity. We provide a descriptive picture of the microstructural factors that discriminate leaders from laggers. We find an intraday profile of lead/lag that reacts to news and market openings. We also study how this lead/lag phenomenon evolves when we only take into account extreme events. We backtest forecasting devices using these lead/lag relationships and find that they are statistically successful, with an average accuracy of about 60% for forecasting variations of the midquote. Interestingly, these lead/lag relationships tend to disappear as we move to larger time scales.

---

*Natixis, Equity Markets. E-mail: nicolas.huth@natixis.com. The authors would like to thank the members of Natixis' statistical arbitrage R&D team for fruitful discussions. We also thank Emmanuel Bacry for his helpful comments.

†BNP Paribas Chair of Quantitative Finance, Ecole Centrale Paris, MAS Laboratory. E-mail: frederic.abergel@ecp.fr

[1] An arbitrage opportunity occurs when one can set up a zero-cost portfolio allowing for strictly positive wealth in the future with non-zero probability.

[2] The Reuters Instrument Code (RIC) of each financial asset is indicated in brackets.

[3] The data used for figure 1 are adjusted closing prices and can be downloaded at fr.finance.yahoo.com for free.

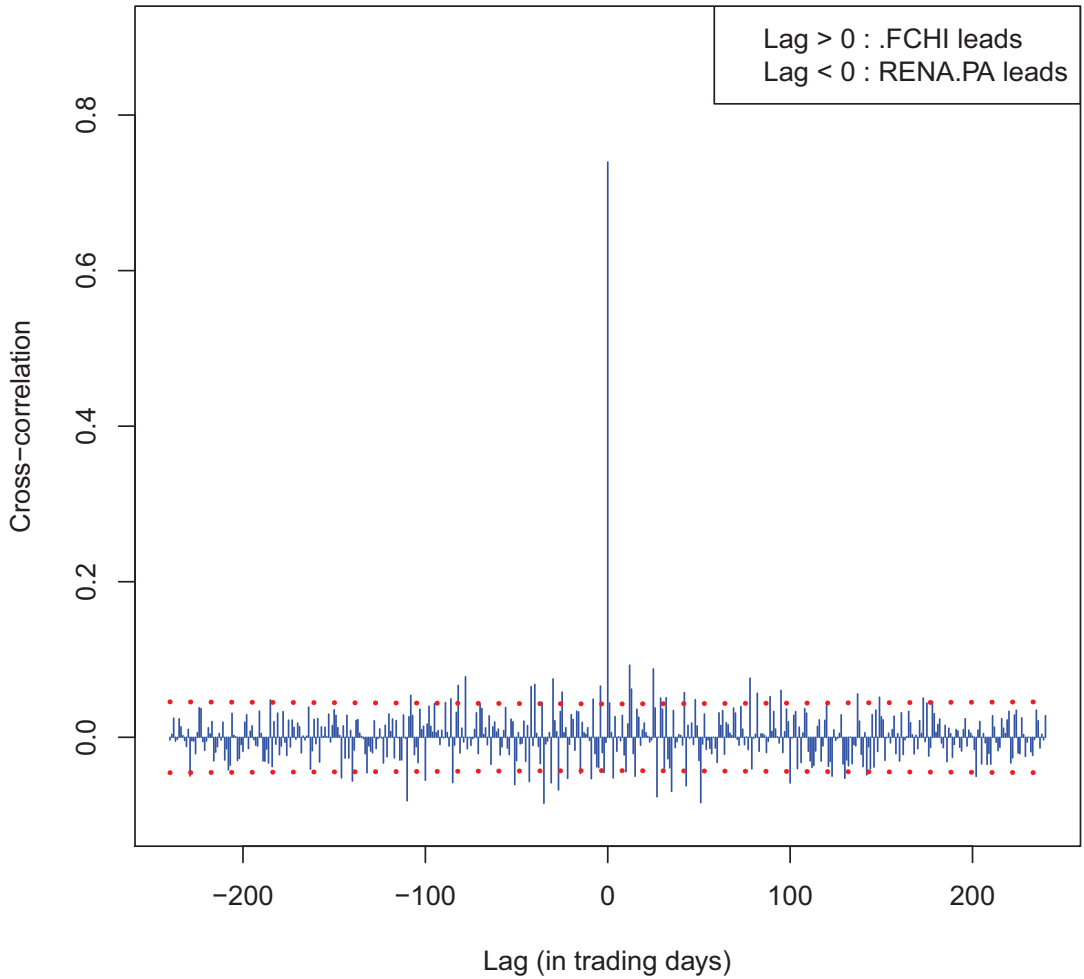

Figure 1: Cross-correlation function between .FCHI and RENA.PA, 2003/01/02 − 2011/03/04

This paper is organized as follows. Section 1 introduces the dataset and provides basic but insightful statistics on the assets under focus. Section 2 describes the methodology used to measure lead/lag relationships. Section 3 presents our empirical results. Finally, section 4 concludes by summarizing the main findings and giving the directions for further research.

# 1 Data description and summary statistics

We have access to the Reuters Tick Capture Engine (RTCE) database which provides tick-by-tick data on many financial assets (equities, fixed income, forex, futures, commodities, etc). Three levels of data are available:

- trades files: each transaction price and quantity timestamped up to the millisecond

- quotes files: each quote (best bid and ask) price or quantity modification timestamped up to the millisecond

- order book files: each limit price or quantity modification timestamped up to the millisecond, up to a given depth, typically ten limits on each side of the order book.

Throughout this study, we will only use trades and quotes files. We will sample quotes on a trading time basis so that for each trade we have access to the quotes right before this trade. Quotes files will also be used to monitor the evolution of quotes in continuous time, for instance to get the quotes right after a trade.

When a trade walks the order book up or down by hitting consecutive limit prices, it is recorded as a sequence of trades with the same timestamp but with prices and quantities corresponding to each limit hit. For instance, assume that the best ask offers 100 shares at price 10 at 200 shares at price 10.01, and that a buy trade arrives for 150 shares. This is recorded as two lines in the trades file with the same timestamp, the first line being 100 shares at 10 and the second line 50 shares at 10.01. As a pre-processing step, we aggregate identical timestamps in the trades files by replacing the price by the volume weighted average price (VWAP) over the whole transaction and the volume by the sum of all quantities consumed. In the previous example, the trade price will thus be $(100*10+50*10.01)/(100+50) = 10.00333$ and the trade quantity $100+50 = 150$.

Table 1 describes the scope of assets for our empirical study. We only consider equities and equity futures. The futures are nearby-maturity futures and are rolled the day before the expiration date. The time period is 2010/03/01-2010/05/31 and the trading hours are specified in table 1. On each day, we drop the first and last half hours of trading. We only consider regular trades to avoid outliers such as block trades or OTC trades (see [4] for a more detailed description). When studying lead/lag relationships between assets traded on different exchanges, we only consider hours of simultaneous trading.

Table 2 gives some insight into the liquidity of each of these assets. It displays the following summary statistics[4]:

- the average duration between two consecutive trades $\langle \Delta t \rangle$

- the average tick size $\delta$ in percentage of the midquote $\langle \delta/m \rangle$

- the average bid/ask spread expressed in tick size $\langle s \rangle / \delta$

- the frequency of unit bid/ask spread $\langle \mathbb{1}_{\{s=\delta\}} \rangle$

- the frequency of trades hitting more than the best limit price available[5] $\langle \mathbb{1}_{\{\text{trade through}\}} \rangle$[4]

- a proxy for the daily volatility expressed in tick size : $\langle |\Delta m| \rangle / \delta$, where $\Delta m$ is the midquote variation between two consecutive trades

- the average turnover per trade $\langle P_{\text{trade}} V_{\text{trade}} \rangle$

Every average is computed independently on a daily basis, and then averaged over all days available :
$\langle x \rangle = \frac{1}{n_{\text{days}}} \sum_{d=1}^{n_{\text{days}}} \frac{\sum_{i=1}^{n_{i,d}} x_{i,d}}{n_{i,d}}$.

---

[4]For assets traded in an other currency than EUR (VOD.L, FFI, NESN.VX and FSMI), we convert the average turnover per trade in EUR by using the closing price of the corresponding exchange rate (GBP/EUR for the two first and CHF/EUR for the two last).

[5]In our data, we detect a trade through as a sequence of trades with the same timestamp and at least two different consecutive execution prices.

Table 1: Description of the scope of assets.

| RIC | Description | Exchange | Trading hours (CET) | Currency |
|---|---|---|---|---|
| ACCP.PA | Accor | Euronext Paris | 09:00-17:30 | EUR |
| AIRP.PA | Air Liquide | Euronext Paris | 09:00-17:30 | EUR |
| ALSO.PA | Alstom | Euronext Paris | 09:00-17:30 | EUR |
| ALUA.PA | Alcatel Lucent | Euronext Paris | 09:00-17:30 | EUR |
| AXAF.PA | Axa | Euronext Paris | 09:00-17:30 | EUR |
| BNPP.PA | BNP Paribas | Euronext Paris | 09:00-17:30 | EUR |
| BOUY.PA | Bouygues | Euronext Paris | 09:00-17:30 | EUR |
| CAGR.PA | Crédit Agricole | Euronext Paris | 09:00-17:30 | EUR |
| CAPP.PA | Cap Gemini | Euronext Paris | 09:00-17:30 | EUR |
| CARR.PA | Carrefour | Euronext Paris | 09:00-17:30 | EUR |
| DANO.PA | Danone | Euronext Paris | 09:00-17:30 | EUR |
| DEXI.BR | Dexia | Euronext Brussels | 09:00-17:30 | EUR |
| EAD.PA | EADS | Euronext Paris | 09:00-17:30 | EUR |
| EDF.PA | EDF | Euronext Paris | 09:00-17:30 | EUR |
| ESSI.PA | Essilor | Euronext Paris | 09:00-17:30 | EUR |
| FTE.PA | France Télécom | Euronext Paris | 09:00-17:30 | EUR |
| GSZ.PA | GDF Suez | Euronext Paris | 09:00-17:30 | EUR |
| ISPA.AS | Arcelor Mittal | Euronext Amsterdam | 09:00-17:30 | EUR |
| LAFP.PA | Lafarge | Euronext Paris | 09:00-17:30 | EUR |
| LAGA.PA | Lagardère | Euronext Paris | 09:00-17:30 | EUR |
| LVMH.PA | LVMH | Euronext Paris | 09:00-17:30 | EUR |
| MICP.PA | Michelin | Euronext Paris | 09:00-17:30 | EUR |
| OREP.PA | L'Oréal | Euronext Paris | 09:00-17:30 | EUR |
| PERP.PA | Pernod Ricard | Euronext Paris | 09:00-17:30 | EUR |
| PEUP.PA | Peugeot | Euronext Paris | 09:00-17:30 | EUR |
| PRTP.PA | PPR | Euronext Paris | 09:00-17:30 | EUR |
| RENA.PA | Renault | Euronext Paris | 09:00-17:30 | EUR |
| SASY.PA | Sanofi Aventis | Euronext Paris | 09:00-17:30 | EUR |
| SCHN.PA | Schneider Electric | Euronext Paris | 09:00-17:30 | EUR |
| SEVI.PA | Suez Environnement | Euronext Paris | 09:00-17:30 | EUR |
| SGEF.PA | Vinci | Euronext Paris | 09:00-17:30 | EUR |
| SGOB.PA | Saint-Gobain | Euronext Paris | 09:00-17:30 | EUR |
| SOGN.PA | Société Générale | Euronext Paris | 09:00-17:30 | EUR |
| STM.PA | StMicroelectronics | Euronext Paris | 09:00-17:30 | EUR |
| TECF.PA | Technip | Euronext Paris | 09:00-17:30 | EUR |
| TOTF.PA | Total | Euronext Paris | 09:00-17:30 | EUR |
| UNBP.PA | Unibail-Rodamco | Euronext Paris | 09:00-17:30 | EUR |
| VIE.PA | Veolia Environnement | Euronext Paris | 09:00-17:30 | EUR |
| VIV.PA | Vivendi | Euronext Paris | 09:00-17:30 | EUR |
| VLLP.PA | Vallourec | Euronext Paris | 09:00-17:30 | EUR |
| VOD.L | Vodafone | London Stock Exchange | 09:00-17:30 | GBP |
| NESN.VX | Nestlé | SIX Swiss Exchange | 09:00-17:30 | CHF |
| DTEGn.DE | Deutsche Telekom | XETRA | 09:00-17:30 | EUR |
| FCE | CAC40 future | NYSE Liffe Paris | 08:00-22:00 | EUR |
| FFI | Footsie100 future | NYSE Liffe London | 09:00-22:00 | GBP |
| FSMI | SMI future | Eurex | 07:50-22:00 | CHF |
| FDX | DAX future | Eurex | 07:50-22:00 | EUR |

Table 2: Summary statistics on the scope of assets.

| RIC | $\langle \Delta t \rangle$ (sec) | $\langle \delta/m \rangle$ (bp) | $\langle s \rangle / \delta$ | $\langle \mathbb{1}_{\{s=\delta\}} \rangle$ (%) | $\langle \mathbb{1}_{\{\text{trade through}\}} \rangle$ (%) | $\langle |\Delta m| \rangle / \delta$ | $\langle P_{\text{trade}} V_{\text{trade}} \rangle$ (EUR$\times 10^3$) |
|---|---|---|---|---|---|---|---|
| ACCP.PA | 13.351 | 1.22 | 3.98 | 16 | 5 | 1.18 | 11 |
| AIRP.PA | 7.327 | 1.15 | 2.76 | 16 | 5 | 0.88 | 13 |
| ALSO.PA | 6.619 | 1.11 | 3.48 | 19 | 6 | 1.01 | 13 |
| ALUA.PA | 8.274 | 4.34 | 1.55 | 58 | 4 | 0.37 | 12 |
| AXAF.PA | 5.578 | 3.3 | 1.37 | 69 | 3 | 0.38 | 17 |
| BNPP.PA | 3.315 | 1.59 | 2.13 | 47 | 6 | 0.65 | 19 |
| BOUY.PA | 9.584 | 1.36 | 2.77 | 27 | 5 | 0.99 | 12 |
| CAGR.PA | 5.908 | 3.35 | 2.34 | 59 | 4 | 0.69 | 13 |
| CAPP.PA | 10.387 | 1.35 | 3.23 | 21 | 5 | 1.01 | 12 |
| CARR.PA | 7.667 | 1.39 | 2.31 | 34 | 4 | 0.76 | 16 |
| DANO.PA | 6.157 | 1.15 | 2.35 | 35 | 5 | 0.7 | 15 |
| DEXI.BR | 22.092 | 2.44 | 5.29 | 8 | 8 | 1.35 | 6 |
| EAD.PA | 12.152 | 3.34 | 1.73 | 50 | 3 | 0.44 | 12 |
| EDF.PA | 8.022 | 1.29 | 2.61 | 30 | 4 | 0.78 | 12 |
| ESSI.PA | 14.45 | 1.08 | 2.83 | 26 | 4 | 0.76 | 10 |
| FTE.PA | 6.579 | 2.97 | 1.18 | 83 | 2 | 0.23 | 20 |
| GSZ.PA | 5.425 | 1.85 | 1.76 | 50 | 4 | 0.47 | 15 |
| ISPA.AS | 3.01 | 1.68 | 2.03 | 39 | 6 | 0.62 | 22 |
| LAFP.PA | 9.033 | 1.6 | 3.17 | 26 | 5 | 0.98 | 14 |
| LAGA.PA | 17.001 | 1.74 | 2.95 | 22 | 4 | 0.86 | 8 |
| LVMH.PA | 6.101 | 1.16 | 2.64 | 19 | 6 | 0.88 | 16 |
| MICP.PA | 9.105 | 1.84 | 2.58 | 28 | 4 | 0.77 | 13 |
| OREP.PA | 10.048 | 1.28 | 2.66 | 19 | 5 | 0.84 | 16 |
| PERP.PA | 12.211 | 1.61 | 2.18 | 35 | 3 | 0.65 | 13 |
| PEUP.PA | 10.107 | 2.36 | 2.81 | 22 | 5 | 0.73 | 12 |
| PRTP.PA | 13.258 | 2.39 | 3.49 | 33 | 4 | 0.92 | 17 |
| RENA.PA | 5.794 | 1.51 | 3.16 | 21 | 6 | 0.99 | 14 |
| SASY.PA | 5.269 | 1.71 | 1.96 | 47 | 4 | 0.49 | 20 |
| SCHN.PA | 6.708 | 1.19 | 2.95 | 15 | 5 | 0.94 | 15 |
| SEVI.PA | 21.398 | 3.1 | 1.82 | 45 | 3 | 0.48 | 8 |
| SGEF.PA | 5.58 | 1.22 | 2.64 | 28 | 5 | 0.81 | 13 |
| SGOB.PA | 6.429 | 1.41 | 2.85 | 24 | 5 | 0.92 | 14 |
| SOGN.PA | 3.351 | 1.2 | 2.91 | 28 | 7 | 0.95 | 15 |
| STM.PA | 13.46 | 1.44 | 3.65 | 10 | 6 | 1.1 | 10 |
| TECF.PA | 12.488 | 1.7 | 4.17 | 12 | 5 | 0.99 | 13 |
| TOTF.PA | 3.283 | 1.21 | 1.88 | 48 | 5 | 0.57 | 22 |
| UNBP.PA | 14.968 | 3.53 | 1.39 | 68 | 2 | 0.35 | 20 |
| VIE.PA | 9.903 | 2.1 | 1.98 | 43 | 3 | 0.52 | 13 |
| VIV.PA | 7.861 | 2.63 | 1.36 | 70 | 3 | 0.35 | 18 |
| VLLP.PA | 11.353 | 3.36 | 1.68 | 52 | 3 | 0.47 | 17 |
| VOD.L | 6.766 | 3.47 | 1.13 | 88 | 2 | 0.26 | 21 |
| NESN.VX | 12.452 | 9.42 | 1.01 | 99 | 1 | 0.05 | 70 |
| DTEGn.DE | 7.315 | 1.56 | 2.58 | 19 | 7 | 0.7 | 29 |
| FCE0 | 1.803 | 1.32 | 1.21 | 83 | 3 | 0.4 | 14 |
| FFI0 | 1.89 | 0.91 | 1.33 | 75 | 3 | 0.47 | 21 |
| FSMI0 | 5.244 | 1.5 | 1.15 | 85 | 3 | 0.38 | 18 |
| FDX0 | 1.215 | 0.83 | 1.28 | 74 | 11 | 0.38 | 24 |

# 2  Methodology

## 2.1  The Hayashi-Yoshida cross-correlation function

In [12], the authors introduce a new[6] estimator of the linear correlation coefficient between two asynchronous diffusive processes. Given two Itô processes $X, Y$ such that

$$dX_t = \mu_t^X dt + \sigma_t^X dW_t^X$$
$$dY_t = \mu_t^Y dt + \sigma_t^Y dW_t^Y$$
$$d\langle W^X, W^Y \rangle_t = \rho_t dt$$

and observation times $0 = t_0 \leq t_1 \leq \ldots \leq t_{n-1} \leq t_n = T$ for $X$ and $0 = s_0 \leq s_1 \leq \ldots \leq s_{m-1} \leq s_m = T$ for $Y$, which must be independent from $X$ and $Y$, they show that the following quantity

$$\sum_{i,j} r_i^X r_j^Y \mathbb{1}_{\{O_{ij} \neq \emptyset\}}$$
$$O_{ij} = ]t_{i-1}, t_i] \cap ]s_{j-1}, s_j]$$
$$r_i^X = X_{t_i} - X_{t_{i-1}}$$
$$r_j^Y = Y_{s_j} - Y_{s_{j-1}}$$

is an unbiased and consistent estimator of $\int_0^T \sigma_t^X \sigma_t^Y \rho_t dt$ as the largest mesh size goes to zero, as opposed to the standard previous-tick correlation estimator[11, 23]. In practice, it amounts to sum every product of increments as soon as they share any overlap of time. In the case of constant volatilities and correlation, it provides a consistent estimator for the correlation

$$\hat{\rho} = \frac{\sum_{i,j} r_i^X r_j^Y \mathbb{1}_{\{O_{ij} \neq \emptyset\}}}{\sqrt{\sum_i (r_i^X)^2 \sum_j (r_j^Y)^2}}$$

Recently, in [13], the authors generalize this estimator to the whole cross-correlation function. They use a lagged version of the original estimator

$$\hat{\rho}(\ell) = \frac{\sum_{i,j} r_i^X r_j^Y \mathbb{1}_{\{O_{ij}^\ell \neq \emptyset\}}}{\sqrt{\sum_i (r_i^X)^2 \sum_j (r_j^Y)^2}}$$
$$O_{ij}^\ell = ]t_{i-1}, t_i] \cap ]s_{j-1} - \ell, s_j - \ell]$$

It can be computed by shifting all the timestamps of $Y$ and then using the Hayashi-Yoshida estimator. They define the lead/lag time as the lag that maximizes $|\hat{\rho}(\ell)|$. In the following we will not estimate the lead/lag time but rather decide if one asset leads the other by measuring the asymmetry of the cross-correlation function between the positive and negative lags. More precisely, we state that $X$ leads $Y$ if $X$ forecasts $Y$ more accurately than $Y$ does for $X$. Formally speaking, $X$ is leading $Y$ if

$$\frac{\|r_t^Y - \text{Proj}(r_t^Y | \vec{r}_{t-}^X)\|}{\|r^Y\|} < \frac{\|r_t^X - \text{Proj}(r_t^X | \vec{r}_{t-}^Y)\|}{\|r^X\|}$$
$$\iff \frac{\|\varepsilon^{YX}\|}{\|r^Y\|} < \frac{\|\varepsilon^{XY}\|}{\|r^X\|}$$

---

[6]In fact, a very similar estimator was already designed in [7].

where $\text{Proj}(r_t^Y | \vec{r}_{t-}^X)$ denotes the projection of $r_t^Y$ on the space spanned by $\vec{r}_{t-}^X := \{r_s^X, s < t\}$. We will only consider the ordinary least squares setting, i.e. $\text{Proj}(r_t^Y | \vec{r}_{t-}^X) = \mu + \int_{]0,\bar{\ell}]} \beta_s r_{t-s}^X ds$ and $\|X\|^2 = \mathbb{V}\text{ar}(X)$. In practice, we compute the cross-correlation function on a discrete grid of lags so that $\int_{]0,\bar{\ell}]} \beta_s r_{t-s}^X ds = \sum_{i=1}^p \beta_i r_{t-\ell_i}^X$. It is easy to show (see appendix A) that

$$\frac{\|\varepsilon^{YX}\|^2}{\|r^Y\|^2} = 1 - (C^{YX})^T (C^{XX} C^{YY})^{-1} C^{YX}$$
$$C^{YX} = (\mathbb{C}\text{ov}(Y_t, X_{t-\ell_1}), \ldots, \mathbb{C}\text{ov}(Y_t, X_{t-\ell_p}))^T$$
$$C^{YY} = \mathbb{V}\text{ar}(Y_t)$$
$$C^{XX} = (\mathbb{C}\text{ov}(X_{t-\ell_i}, X_{t-\ell_j}), i, j = 1, \ldots, p)$$

$(C^{YX})^T (C^{XX} C^{YY})^{-1} C^{YX}$ measures the correlation between $Y$ and $X$. Indeed, $X$ is a good predictor of $Y$ if both are highly correlated. If we assume furthermore that the predictors $X$ are uncorrelated, we can show (see appendix A) that

$$\frac{\|\varepsilon^{YX}\|}{\|r^Y\|} < \frac{\|\varepsilon^{XY}\|}{\|r^X\|}$$
$$\iff \sum_{i=1}^p \rho^2(\ell_i) > \sum_{i=1}^p \rho^2(-\ell_i)$$
$$\iff \text{LLR} := \frac{\sum_{i=1}^p \rho^2(\ell_i)}{\sum_{i=1}^p \rho^2(-\ell_i)} > 1$$

The asymmetry of the cross-correlation function, as defined by the LLR (standing for Lead/Lag Ratio) measures lead/lag relationships. This definition of lead/lag is closely related to the notion of Granger causality[10]. Given two stochastic processes $X$ and $Y$, $X$ is said to cause $Y$ (in the Granger sense) if, in the following linear regression

$$Y_t = c + \sum_{k=1}^p a_k^{YX} \cdot Y_{t-k} + \sum_{\ell=1}^q b_\ell^{YX} \cdot X_{t-\ell} + \varepsilon_t^{YX}$$

some of the estimated coefficients $\hat{b}_\ell^{YX}$ are found to be statistically significant. Since these coefficients are closely linked to the cross-correlation function, there is indeed a strong similarity between this approach and ours. The Granger regression includes lags of the lagger in order to control for its autocorrelation, which we do not take into account. Note that there is *a priori* no obstacle to find that $X$ Granger-causes $Y$ and $Y$ Granger-causes $X$. Our approach amounts to compare in some sense the signficance of all $\hat{b}_\ell^{YX}$ to the one of all $\hat{b}_\ell^{XY}$.

Our indicator tells us which asset is leading the other for a given pair, but we might also wish to consider the strength and the characteristic time of this lead/lag relationship. Therefore, the maximum level of the cross-correlation function and the lag at which it occurs must also be taken into account.

In the following empirical study, we measure the cross-correlation function between variations of midquotes of two assets, i.e. $X$ and $Y$ are midquotes. The observation times will be tick times. Tick time is defined as the clock that increments each time there is a non-zero variation of the midquote between two trades (not necessarily consecutive). It does not take into account the nil variations of the midquote, contrary to trading time. Computing the Hayashi-Yoshida correlation in trade time or in tick time does not yield the same result. Indeed, consider the trading sequence on figure 2. It is easily seen that the trade time covariance is zero while the tick time covariance is not. Since we will be interested in forecasting the midquote variation of the lagging asset, we prefer to use tick time rather than trade time since classification in tick time is binary: either the midquote moves up or it moves down.

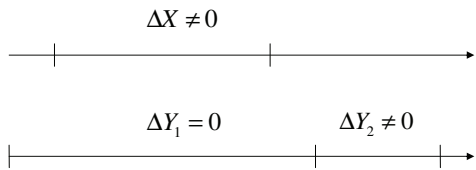

Figure 2: Difference between the trade time and tick time Hayashi-Yoshida correlations

## 2.2 Simulation study: artificial lead/lag due to different levels of trading activity

When looking at two assets, a natural bet is to say that the most traded asset leads the other. In the literature, empirical research focusing on lead/lag relationships[7, 8, 14, 16, 17, 18, 19, 20, 21] often concludes that the most liquid assets drive the others[7]. Intuitively, the most heavily traded assets tend to incorporate information into prices faster than others so they should lead.

A very simple simulation framework[11, 12] provides some insight into this liquidity lead/lag effect. Let us consider two correlated Brownian motions $B_1, B_2$ with correlation $\rho$ on $[0, T]$ and two series of random timestamps $0 = t_0 < t_1 < \ldots < t_n = T$ and $0 = s_0 < s_1 < \ldots < s_m = T$ independent of $B_1, B_2$. For instance, let the timestamps be the jumping times of two independent Poisson processes with respective intensities $\lambda_1$ and $\lambda_2$. We define two time series of price as the Brownian motions sampled along the Poisson timestamps

$$X(u) = B_1(t(u))$$
$$Y(u) = B_2(s(u))$$
$$t(u) = \max\{t_i | t_i \leq u\}$$
$$s(u) = \max\{s_i | s_i \leq u\}$$

There should be no lead/lag relationship between $X$ and $Y$ since they are sampled from two synchronous Brownian motions. The cross-correlation function should thus be a Dirac delta function with level $\rho$ at lag zero. Figure 3 illustrates the behaviour of the cross-correlation function computed with the previous-tick and the Hayashi-Yoshida estimators for various levels of $\frac{\lambda_1}{\lambda_2}$. We simulate two synchronously correlated Brownian motions on $[0, T = 30600]$ with time step $\Delta t = 5$ and correlation $\rho = 0.8$. Then we sample them along two independent Poisson time grids with parameters $\lambda_1$ and $\lambda_2$. We repeat this simulation 64 times and average the cross-correlation functions computed independently over each simulation. In the Hayashi-Yoshida case, we also plot on figure 3 the average cross-correlation function computed using the closed-form formula shown in appendix B.

From figure 3, we observe that the cross-correlation function is always peaked at zero, whatever the method of computation. The previous-tick correlation function goes to zero with the level of asynchrony, but we get rid of this problem with the Hayashi-Yoshida estimator. The previous-tick correlation function is blurred by spurious liquidity effects: the asymmetry grows significantly with $\frac{\lambda_1}{\lambda_2}$, yielding he most active Brownian motion to always lead the other. In the contrary, the Hayashi-Yoshida LLR is not impacted by the level of $\frac{\lambda_1}{\lambda_2}$ and it remains symmetric (see appendix B for the proof). Even though the Hayashi-Yoshida cross-correlation remains symmetric, it is not exactly a Dirac mass. The irregular sampling creates correlation at non-zero lags (see appendix B for the proof).

---

[7] Liquidity does not necessarily mean more transactions, it can be measured with other microstructure statistics such as bid/ask spread, market impact etc...

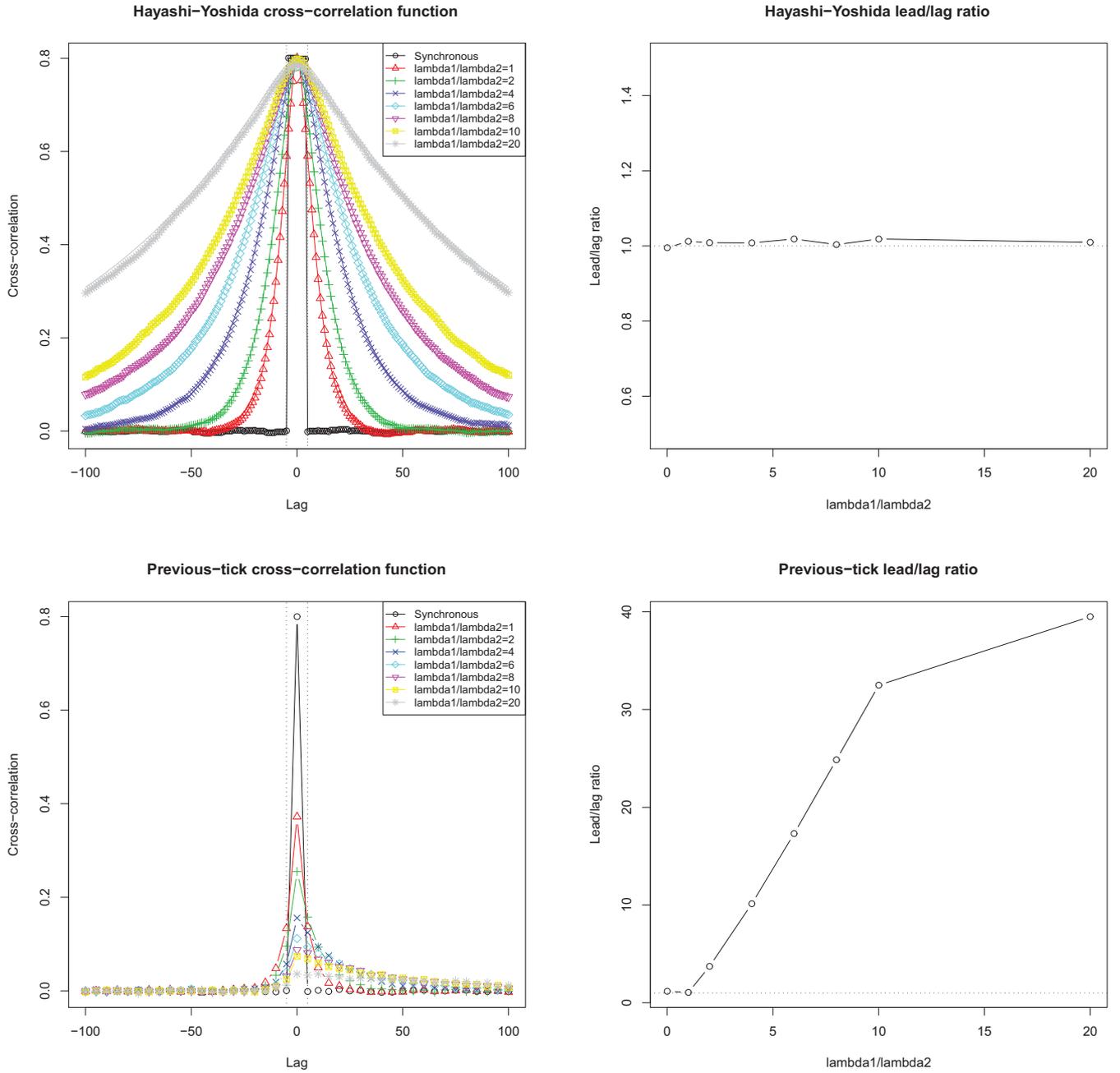

Figure 3: Cross-correlation of two synchronously correlated Brownian motions sampled along Poisson time grids for various levels of $\frac{\lambda_1}{\lambda_2}$, with $\lambda_1 = \frac{1}{\Delta t} = 0.2$ kept fixed. Top left panel: Hayashi-Yoshida cross-correlation function and its closed-form expression (straight lines). Top right panel: Hayashi-Yoshida LLR. Bottom left panel: Previous-tick cross-correlation function. Bottom right panel: Previous-tick LLR.

As a result, we choose to use the Hayashi-Yoshida cross-correlation function to measure lead/lag relationships in our empirical study. This allows us to avoid being fooled by liquidity effects, yielding the most traded assets to be automatically leaders.

## 3 Empirical results

### 3.1 Empirical cross-correlation functions

We now turn to measuring lead/lag relationships on our dataset. Figure 4 shows the tick time Hayashi-Yoshida cross-correlation functions computed on four pairs of assets

- FCE/FSMI: future/future
- FCE/TOTF.PA: future/stock
- RENA.PA/PEUP.PA: stock/stock
- FSMI/NESN.VX: future/stock

We choose the following grid of lags (in seconds)

$$0, 0.01, 0.02, \ldots, 0.1, 0.2, \ldots, 1, 2, \ldots, 10, 15, 20, 30, \ldots, 120, 180, 240, 300$$

We consider that there is no correlation after five minutes of trading on these assets, which seems to be empirically justified on figure 4, except for the FSMI/NESN.VX case. Figure 5 is similar to figure 4, but zooms on lags smaller than 10 seconds. In order to assess the robustness of our empirical results against the null hypothesis of no genuine lead/lag relationship but only artificial liquidity lead/lag, we build a surrogate dataset. For two assets and for a given trading day, we generate two synchronously correlated Brownian motions with the same correlation as the two assets $\rho = \hat{\rho}_{HY}(0)$ on $[0, T]$, $T$ being the duration of a trading day, with a mesh of one second. Then we sample these Brownian motions along the true timestamps of the two assets, so that the surrogate data have the same timestamp structure as the original data. The error bars indicate the 95%-confidence interval for the average correlation over all trading days[8].

For the FCE/FSMI pair (future *vs* future), the cross-correlation vanishes very quickly, there is less than 5% of correlation at 30 seconds. We observe that there is more correlation on the side where FCE leads with a LLR of 1.26 and a maximum correlation at 0.2 seconds. The pair FCE/TOTF.PA involves a future on an index and a stock being part of this index. Not surprisingly, the future leads the stock, by an average time of 0.6 seconds. This pair shows the biggest amount of lead/lag as measured by the LLR (2.12). The RENA.PA/PEUP.PA case compares two stocks in the French automobile industry. The cross-correlation function is the most symmetric of the four shown, with a LLR of 1.15 and an average maximum lag of 0.95 seconds. Note that the maximum lag displays a signifcantly larger standard deviation than for the two previous pairs. It confirms that the lead/lag effect for these two assets is not as much pronounced as for the two pairs considered before. Finally, the FSMI/NESN.VX pair is interesting because the stock leads the future on the index where it belongs. It might be explained by the fact NESN.VX is the largest market capitalization in the SMI, about 25%. The asymmetry is quite strong (LLR = 0.69) and the maximum lag is 17 seconds. However, the standard deviation of the maximum lag is pretty strong, almost half the average maximum lag. We also see that there still exists significant correlation after five minutes. The difference between the maximum correlation and the correlation at lag zero is 6% for FCE/FSMI, 7% for FCE/TOTF.PA, 0.3% for RENA.PA/PEUP.PA and 2% for FSMI/NESN.VX, which confirms again that the lead/lag is less pronounced for RENA.PA/PEUP.PA. The LLR for surrogate data is equal to one and the maximum lag is statistically zero with a usual confidence level of 95% for the four pairs of assets considered. This strong contrast between real and surrogate data suggests that there are genuine lead/lag relationships between these assets that are not solely due to differences in the level of trading activity.

---

[8]Assuming our dataset is made of $D$ uncorrelated trading days, the confidence interval for the average correlation $\bar{\rho}_D = \frac{1}{D}\sum_{d=1}^{D}\rho_d$ is $\left[\bar{\rho}_D \pm 1.96\frac{\sigma_D}{\sqrt{D}}\right]$ where $\sigma_D^2 = \frac{1}{D}\sum_{d=1}^{D}\rho_d^2 - \bar{\rho}_D^2$. By doing so, we neglect the variance of the correlation estimator inside a day.

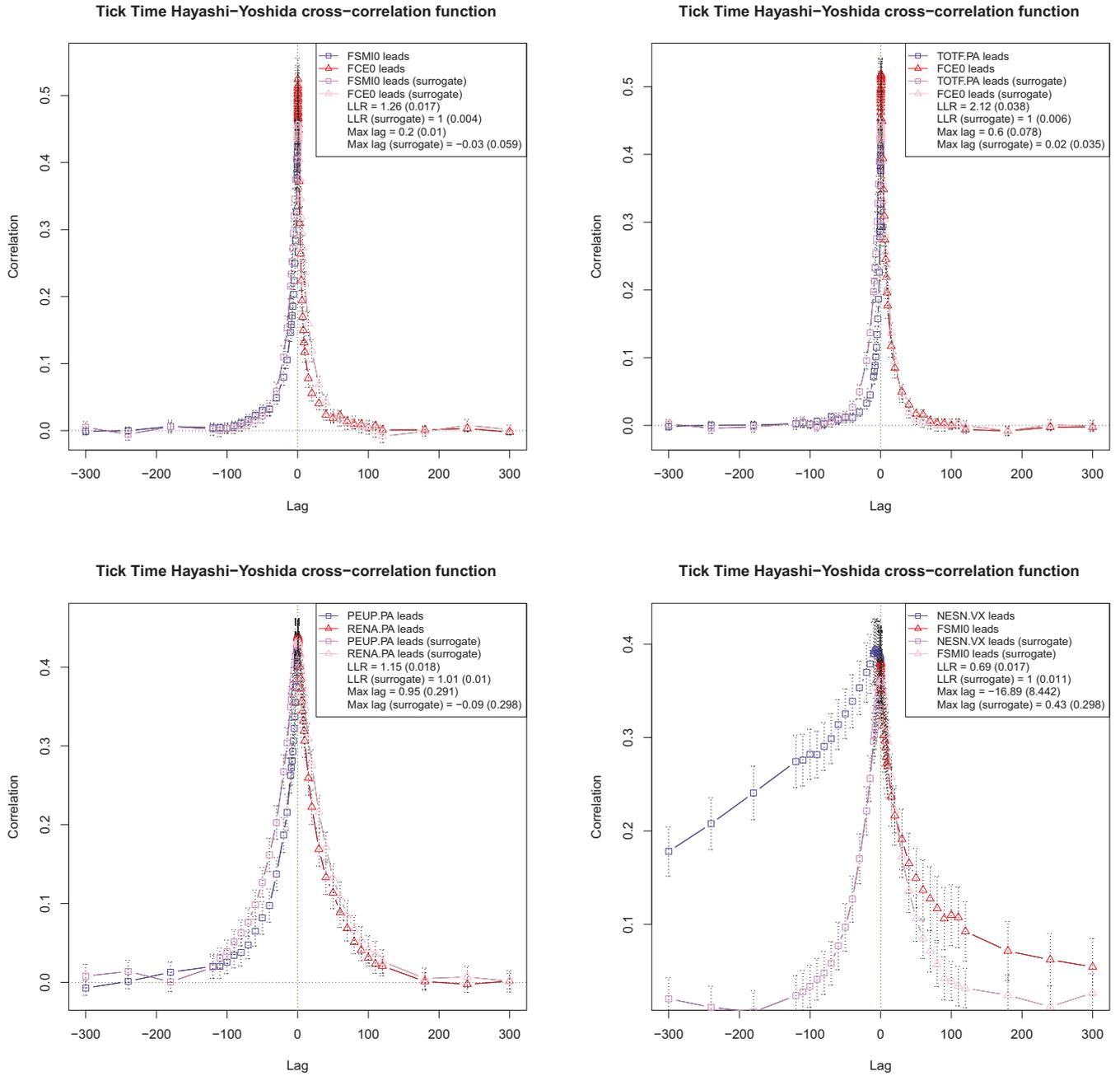

Figure 4: Tick time Hayashi-Yoshida cross-correlation function. Top left panel: FCE/FSMI. Top right panel: FCE/TOTF.PA. Bottom left panel: RENA.PA/PEUP.PA. Bottom right panel: FSMI/NESN.VX. The standard deviations are indicated between brackets.

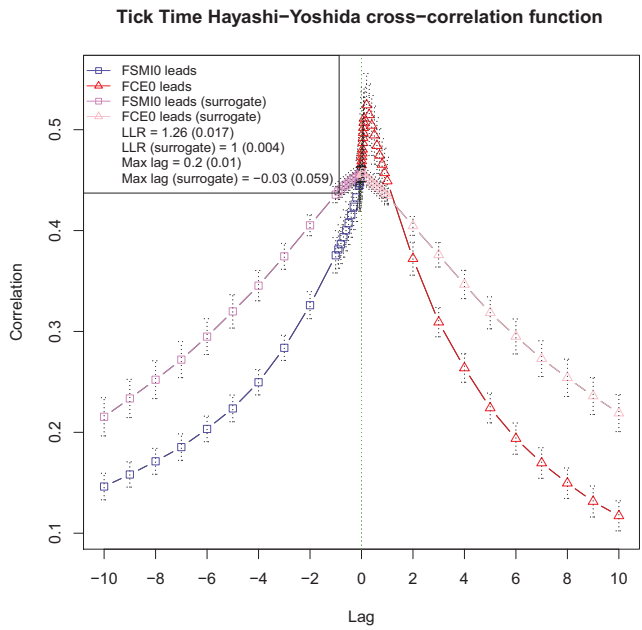
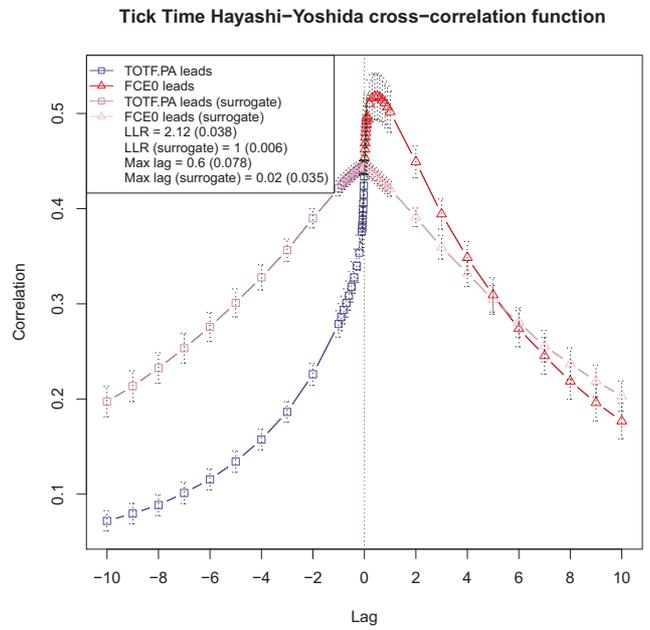
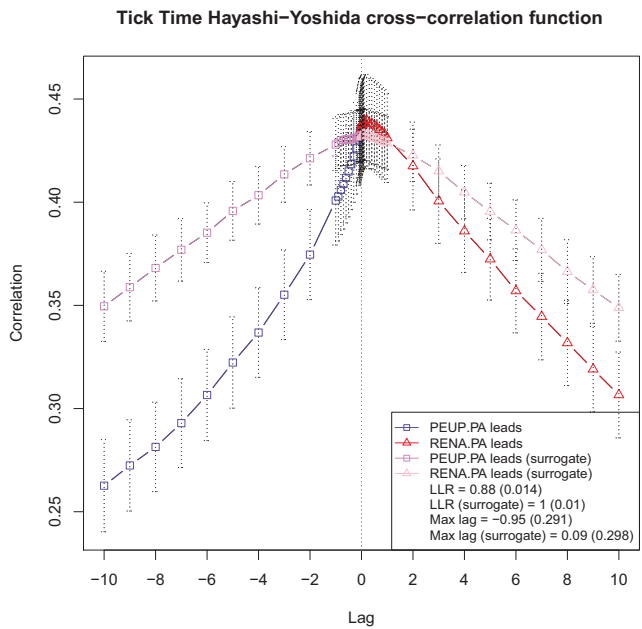
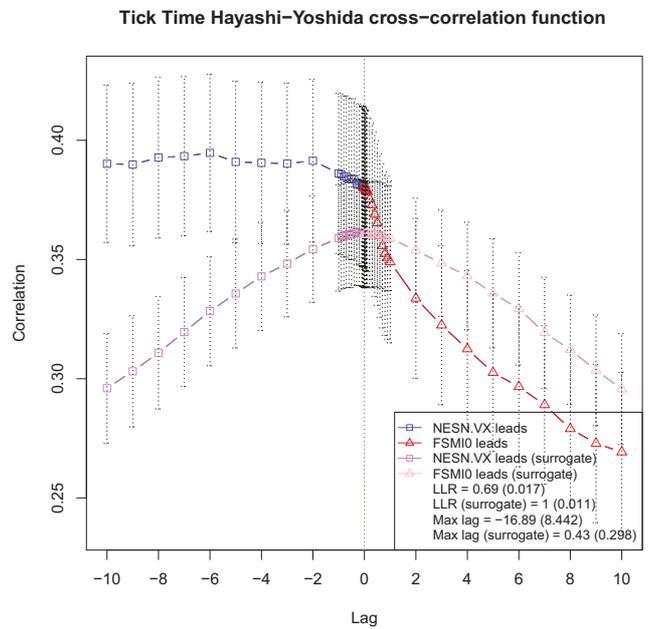

Figure 5: Zoom on lags smaller than 10 seconds in figure 4.

## 3.2 Microstructure features of leading assets

In this section, we investigate what are the common features of leading assets. It is often claimed in the literature [7, 8, 14, 16, 17, 18, 19, 20, 21] that the most liquid assets tend to be leaders, which sounds intuitive because it should take more time to illiquid assets to incorporate information into prices.

In section 1, we have presented several indicators measuring liquidity from the point of view of market microstructure. We now look for the dependency of our lead/lag indicator LLR on these liquidity indicators. In order to make an extensive study, we have computed the LLR and six liquidity indicators[9] for all pairs in the universe made from the CAC40 components and its future, which amounts to $41*40/2 = 820$ pairs. To be more precise, for all pairs $(X, Y)$, and for each indicator I, we plot the LLR against the ratio IR $= \frac{I_X}{I_Y}$. Remembering that the LLR is the ratio of the squared correlations when $X$ leads over those when $Y$ leads, it means that if sign(LLR $-1$) = sign(IR $-1$), then the higher this indicator, the more $X$ leads and *vice versa*. The results are shown in figure 6. Table 3 illustrates the discriminatory power of each of these indicators by counting the proportion of points falling into the four quadrants delimited by the straight lines $x = 1$ and $y = 1$.

Table 3: Discriminatory power of liquidity indicators.

|  | $N^{++}$ | $N^{--}$ | $N^{+-}$ | $N^{-+}$ | $N^{++} + N^{--}$ | $N^{+-} + N^{-+}$ |
|---|---|---|---|---|---|---|
| $\langle \Delta t \rangle$ | 7% | 6% | 44% | 43% | 13% | 87% |
| $\langle \delta/m \rangle$ | 22% | 26% | 28% | 24% | 48% | 52% |
| $\langle \mathbb{1}_{\{\text{trade through}\}} \rangle$ | 29% | 23% | 22% | 26% | 52% | 48% |
| $\langle s \rangle / \delta$ | 17% | 12% | 33% | 38% | 29% | 71% |
| $\langle |\Delta m| \rangle / \delta$ | 20% | 13% | 31% | 36% | 33% | 67% |
| $\langle P_{\text{trade}} \cdot V_{\text{trade}} \rangle$ | 35% | 42% | 16% | 7% | 77% | 23% |

The most discriminatory indicators are the intertrade duration, the average turnover per trade, the average bid/ask spread and the midquote volatility. The tick size and the probability of having a trade through do not seem to play any direct role in determining who leads or lags. The most liquid assets appear to be leaders, which is in agreement with common market knowledge. Indeed, assets which trade faster, or involve bigger exchanges of money, or have a narrower bid/ask spread, or are less volatile tend to lead on average. Even though the number of trade through does not emerge as a key feature at first sight, we see a decreasing trend if we focus on the future/stock pairs, i.e. the blue points on figure 6, which means that stocks having a bigger probability of trade through are less led by the future. This is still in agreement with the intuition that the most liquid assets tend to lead.

On figure 7, we plot the (cross-sectional) average maximum correlation per decile of ratio of liquidity indicators. In other words, we bin pairs of assets according to the cross-sectional distribution of ratio of liquidity indicators, and we compute the average maximum correlation in each bin. Most of the weight of these distributions is concentrated around where ratios of liquidity are close to 1. It means that highly correlated stocks tend to have a similar level of liquidity, as measured by the six indicators above. As a result, there is a trade-off in lead/lag relationships: while liquid *vs* illiquid pairs exhibit highly asymmetric cross-correlation functions, they tend to be less correlated than pairs with similar liquidity.

Figure 8 provides a network of stock/stock lead/lag relationships in the CAC40 universe[18]. We use a minimum spanning tree[5] to plot the network, which only keeps the most significant correlations by construction. We draw a directed edge from stock $X$ to stock $Y$ if stock $X$ leads stock $Y$, i.e. LLR $= \frac{\sum_{\ell>0} \rho^2(\text{X leads Y by } \ell)}{\sum_{\ell>0} \rho^2(\text{Y leads X by } \ell)} > 1$. The color of the edge indicates the level of LLR of the associated pair. This network can be useful to find optimal pairs of assets for lead/lag arbitrage. Indeed, good candidates are close nodes (high correlation) with red links (high LLR).

---

[9]We omit the probability of unit bid/ask spread because it essentially gives the same information as the average spread.

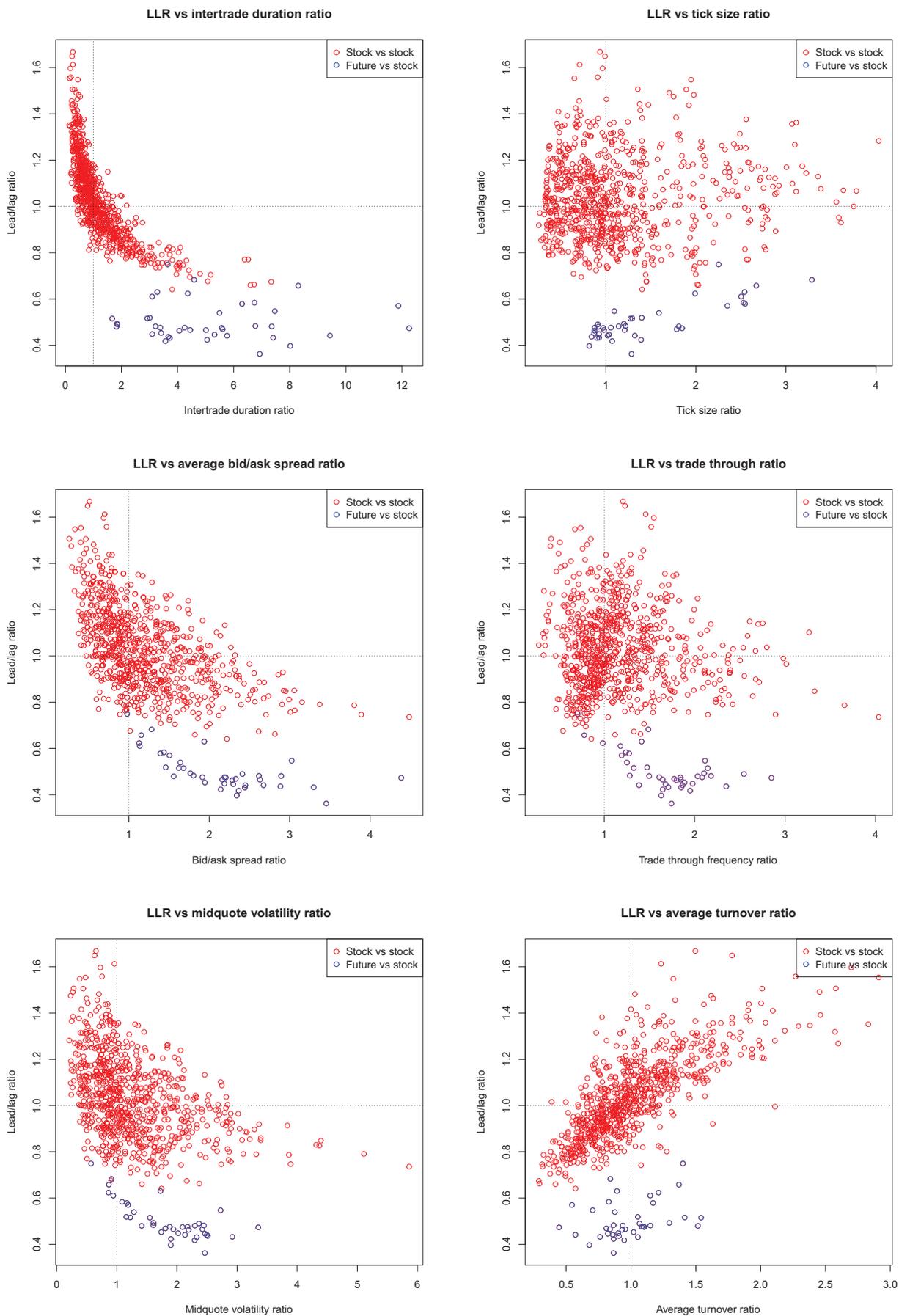

Figure 6: Scatterplot of LLR against pairwise ratios of various liquidity indicators for all the pairs in the CAC40 universe and its future.

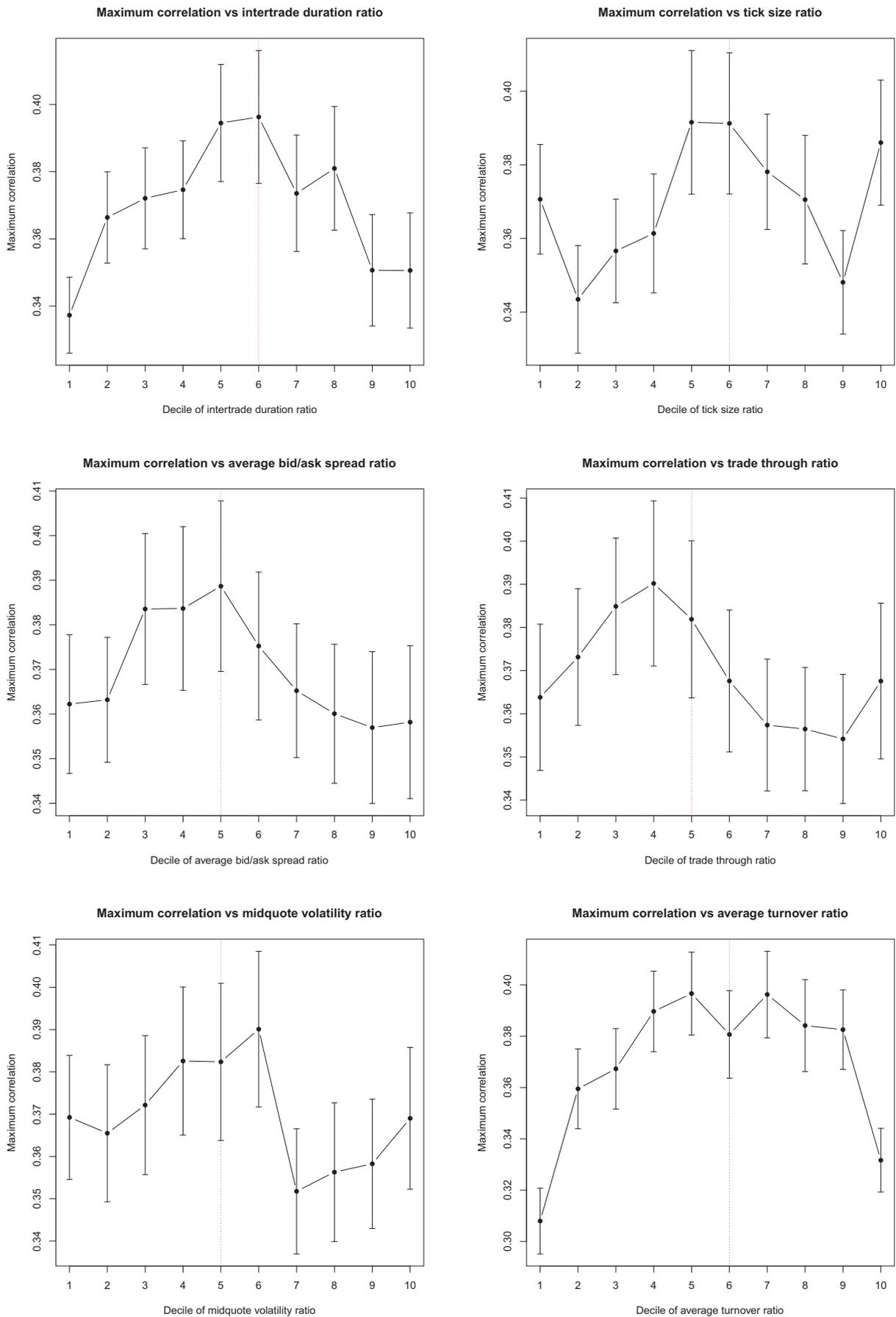

Figure 7: Average maximum correlation against deciles of pairwise ratios of various liquidity indicators for all the pairs in the CAC40 universe and its future. A dotted red line indicates the decile that includes 1. Error bars represent Gaussian 95%-confidence intervals.

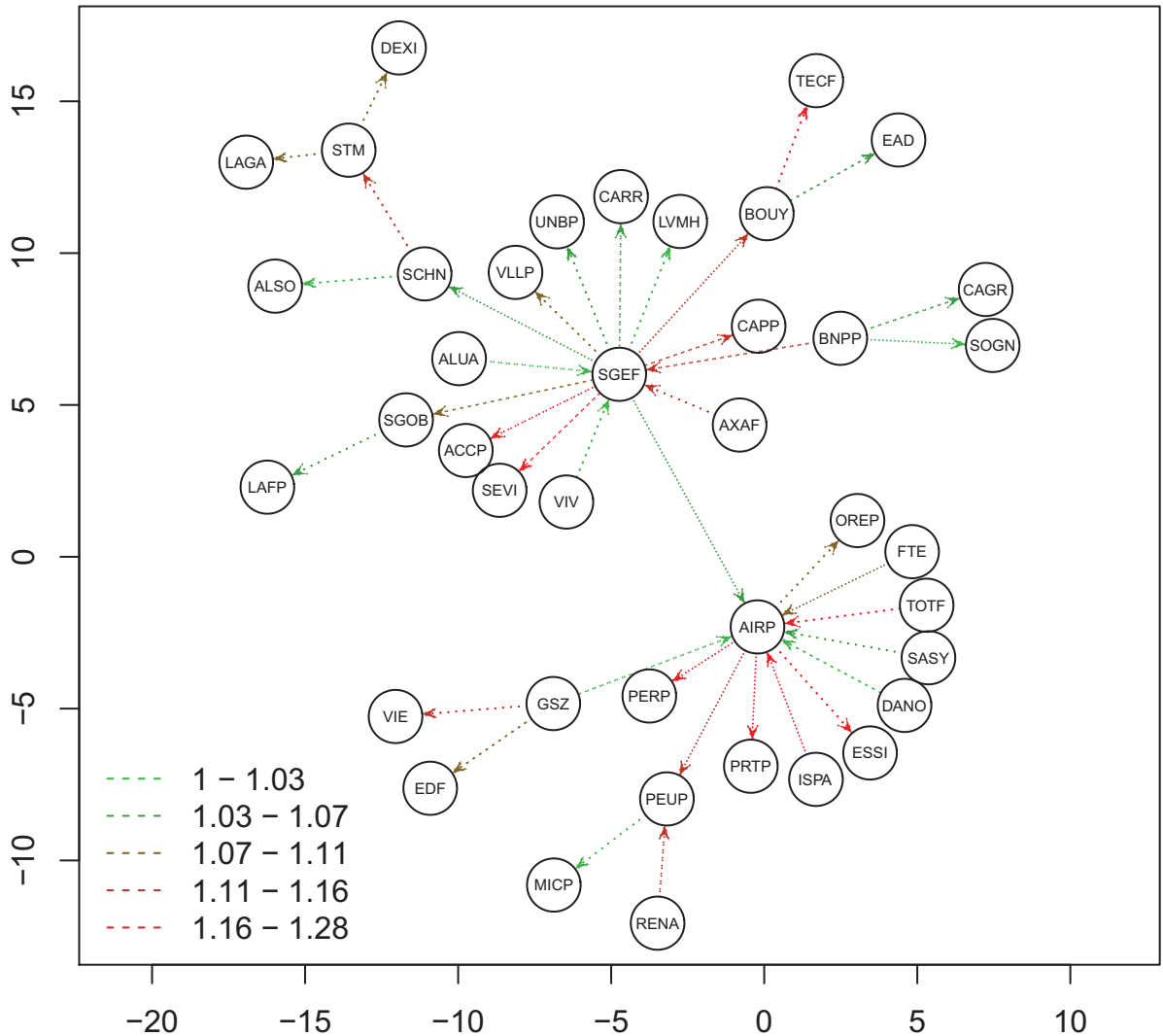

Figure 8: Lead/lag network on the CAC40 universe. The axes are arbitrary.

### 3.3 Intraday profile of lead/lag

A well known stylized fact about financial markets activity is that it strongly changes over the day[3]. For instance, the intraday volatility exhibits a so-called asymmetric U-shape: massive volatility at the open, then it decreases to reach a minimum during lunch time, it peaks at macroeconomic figures announcements, and even experiences a change of regime (in Europe) after the opening of the US market, and finally rallies again at the close, but it does not recover the opening level.

We study the same phenomenon for lead/lag relationships by computing our three lead/lag indicators (LLR, maximum lag measured in seconds and maximum correlation) into 5-minute slices from the open to the close and averaging over all days for each slice[10]. Figure 9 plots the results for future/stock pairs with whisker plots[11] describing the cross-sectional distribution (i.e. the distribution among assets) for each time slice. Figure 10 does the same for stock/stock pairs.

For future/stock pairs, the LLR is always above 1 so that the future leads stocks all day long. The LLR is quite low at the opening and then jumps up five minutes after, it exhibits a noisy U-shape from 09:30 to 14:00; it drops between 14:20 and 14:30, between 15:15 and 15:30, and between 15:45 and 16:00; finally it decreases half an hour before the market closes to reach its lowest level. The maximum lag is always positive most of the time, which confirms that the future always leads stocks. We also notice that lead/lag becomes faster at 14:30 and 16:00 (announcement of US macroeconomic figures) and 15:30 (US market opening), where the maximum lag reaches local minima. There is a global upward trend in the maximum correlation as we move forward on the timeline[3], still with significant peaks at the aforementioned specific event times, and a decorrelation as the market closing approaches.

Stock/stock pairs also show a varying intraday profile of lead/lag. We first remark that lead/lag relationships are far less pronounced than in the future/stock case. Indeed, the LLR is around 1.2 on average, while it is 2.2 for future/stock pairs. The average level of correlation is similar in both cases, though there are seldom uncorrelated future/stock pairs in comparison with stock/stock pairs. Indeed, two stock that belong to very different business sectors might be little correlated, but both are strongly correlated with the future. For instance, the percentage of stock/stock pairs having a correlation less than 0.3 is 19% while it is 7% for future/stock pairs. The LLR is at its highest level at the open, which might reflect the fact that some corporate news are discovered when the market is closed, then it decreases until 10:00 and stays constant until 17:00 after which it drops until the close at its lowest level. The decay of the maximum lag is similar to the one observed for LLR and shows that stock/stock cross-correlation functions tend to be symmetric around zero as time goes by. Finally, the maximum correlation shows the same rising profile than for future/stock pairs. This comes from the fact that most of the correlation comes from the so-called "market mode"[1]. However, it is reported that stock-specific correlations (i.e. once the market mode is statistically removed) tend to decrease during the day[3].

---

[10]More precisely, since we don't have so many data points during 5 minutes, we only consider lags no larger than a minute and we rather compute the cross-correlation function for each day and each slice and we average these cross-correlation functions over days, which gives us one cross-correlation function per slice. We also interpolate these cross-correlation functions with a spline on a regular grid of lags with mesh 0.1 second (function `spline` of R). Then we compute the maximum lag and maximum correlation with these smooth cross-correlation functions, but the LLR is computed using only values on the non-interpolated grid to make it comparable with values obtained in the previous sections. The same approach is used in section 3.4 but we consider lags up to 300 seconds.

[11]The whisher plots we present display a box ranging from the first to the third quartile with the median in the middle, and whiskers extending to the most extreme point that is no more than 1.5 times the interquartile range from the box. These are the default settings in the `boxplot` function of R.

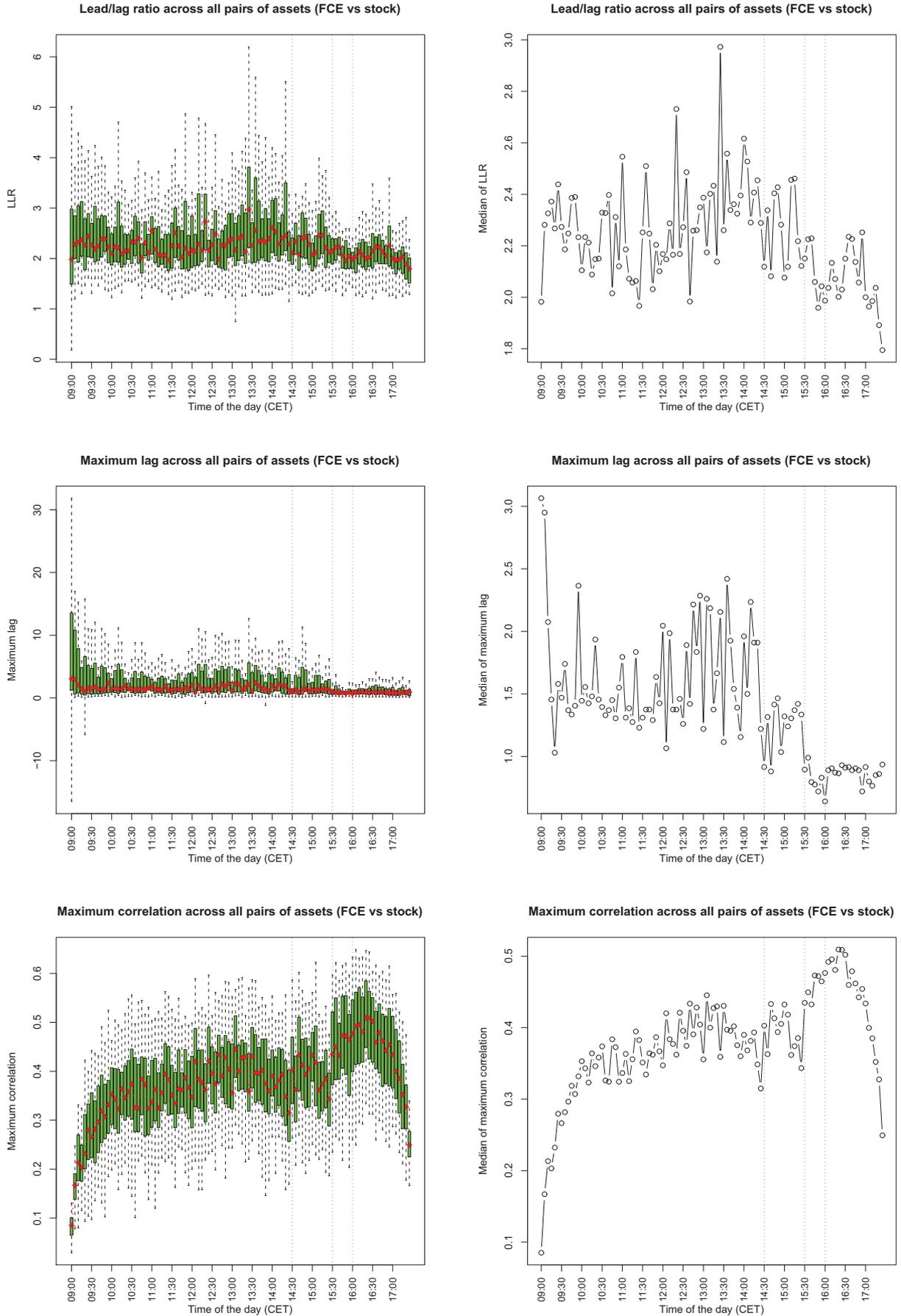

Figure 9: Intraday profile of lead/lag for future/stock pairs. Left panel: cross-sectional distribution for LLR, maximum lag and maximum correlation. Right panel: zoom on cross-sectional medians. Blue dotted lines are drawn 14:30 and 16:00 (announcement of US macroeconomic figures) and 15:30 (NYSE and NASDAQ opening).

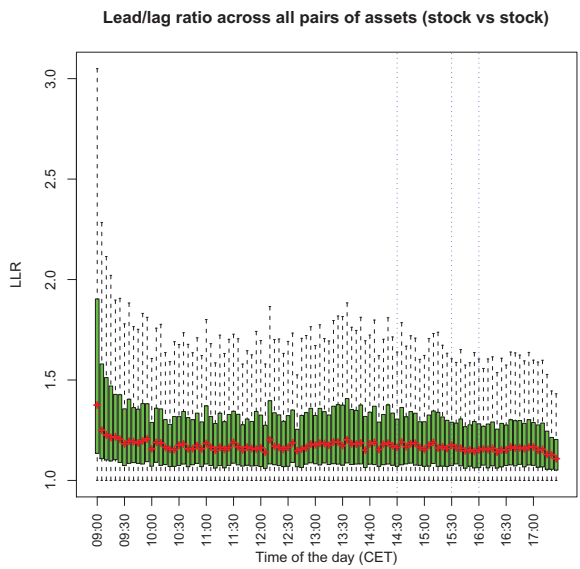
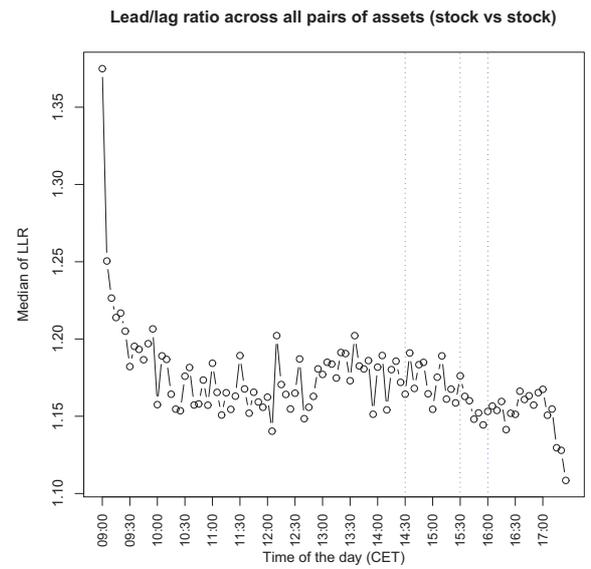
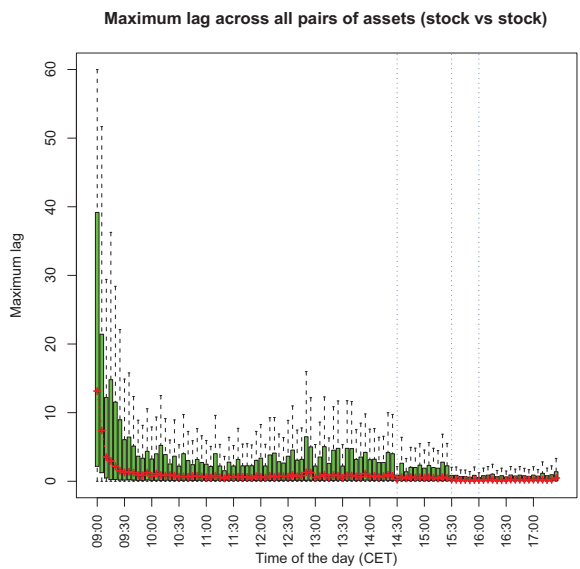
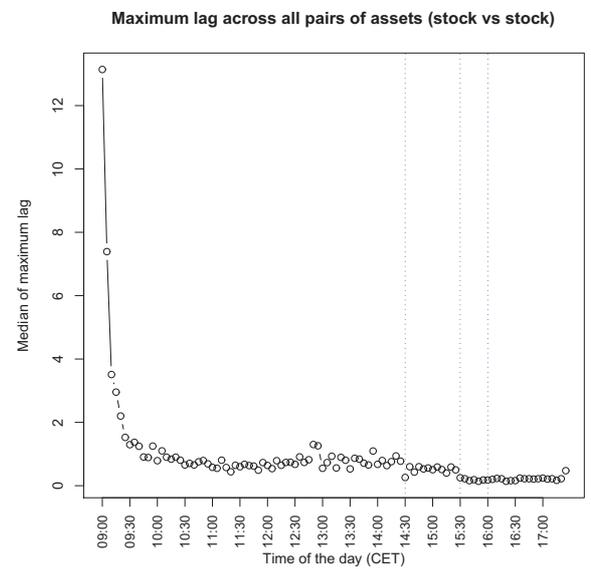
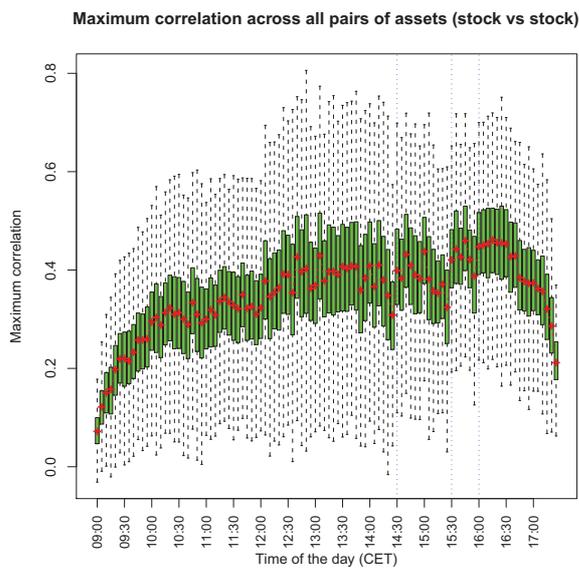
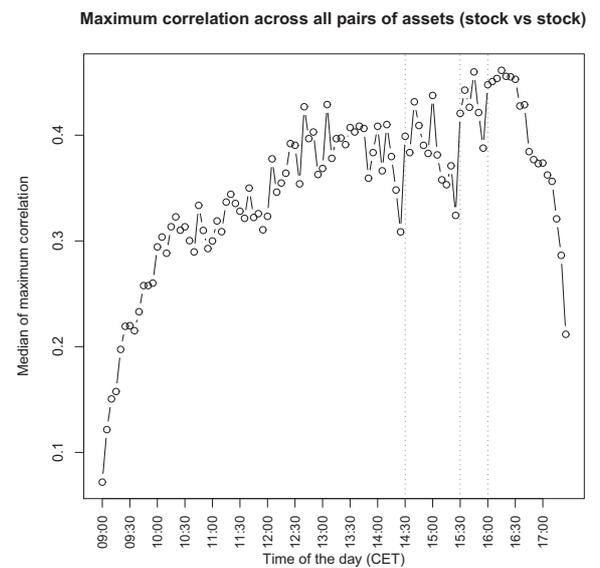

Figure 10: Same as figure 7 for stock/stock pairs.

## 3.4 Lead/lag conditional to extreme events

In the previous sections, we measured lead/lag relationships taking into account every non-zero price variation, whatever its magnitude. However, it sounds reasonable that large returns are more informative than small ones[4]. We introduce a thresholded version of the cross-correlation estimator

$$\hat{\rho}_\theta(\ell > 0) = \frac{\sqrt{N_\theta^X N_0^Y}}{N_\theta^{X,Y}(\ell)} \frac{\sum_{i,j} r_i^X r_j^Y \mathbb{1}_{\{O_{ij}^\ell \neq \emptyset\}} \mathbb{1}_{\{|r_i^X| \geq \theta\}}}{\sigma_\theta^X \sigma_0^Y}$$

$$\hat{\rho}_\theta(\ell < 0) = \frac{\sqrt{N_0^X N_\theta^Y}}{N_\theta^{Y,X}(\ell)} \frac{\sum_{i,j} r_i^X r_j^Y \mathbb{1}_{\{O_{ij}^\ell \neq \emptyset\}} \mathbb{1}_{\{|r_j^Y| \geq \theta\}}}{\sigma_0^X \sigma_\theta^Y}$$

$$\sigma_\theta^k = \sqrt{\sum_i (r_i^k)^2 \mathbb{1}_{\{|r_i^k| \geq \theta\}}}$$

$$N_\theta^k = \sum_i \mathbb{1}_{\{|r_i^k| \geq \theta\}}$$

$$N_\theta^{k,p}(\ell) = \sum_{i,j} \mathbb{1}_{\{|r_i^k| \geq \theta\}} \mathbb{1}_{\{|r_j^p| \geq 0\}} \mathbb{1}_{\{O_{ij}^\ell \neq \emptyset\}}$$

We only take into account price variations of the leading asset that are greater or equal than some threshold $\theta$. For $\ell = 0$, we compute both possibilities. Figure 11 plots the LLR, maximum lag and maximum correlation as a function of $\theta = i * \delta/2, i = 1, \ldots, 6$ for future/stock and stock/stock pairs[12].

The overall trend is that lead/lag becomes more and more pronounced as we focus on larger price variations. Indeed, both the LLR and the maximum correlation increase with the threshold $\theta$. There is roughly two times more correlation when $\theta$ goes from 0.5 to 3. The maximum lag is quite independent from $\theta$. Figure 11 suggests that one should filter out insignificant moves of the leader when trying to build up a forecast of the lagger.

## 3.5 Lead/lag response functions

The cross-correlation gives a lot of insight in the detection of lead/lag relationships. However, it does not tell us the strength of the variation of the lagger following a variation of the midquote of the leader. From a practical point of view, it is of great importance because an arbitrage strategy based on market orders is only profitable if it generates enough profit to bypass the bid/ask spread. As a result, we study the so-called lead/lag response functions

$$R_{v, \lessgtr}(\ell, \theta) = \left\langle v_{t+\ell}^{\text{stock}} - v_t^{\text{stock}} | r_t^{\text{future}} \lessgtr \theta \right\rangle$$

for $v$ being any relevant variable in the order book, such as the bid and ask quotes or the bid/ask spread. The main issue in measuring such a function is that after a jump of the future, one can only record the trajectory of the stock before any another jump of the future happens if we want to isolate the impact of that particular jump. Since futures are much more actively traded than stocks, this can lead to a substantial lack data for large lags, which is why we show the results for lags less than 10 seconds. We need to monitor the state of the best quotes of the stock continuously so we use the quotes files (see section 1).

Figure 12 plots the response function for FCE/TOTF.PA and for $v$ being the bid/ask quotes and the bid/ask spread. The same graphs for FDX/DTEGn.DE, FFI/VOD.L and FSMI/NESN.VX are displayed in appendix C. The first row of figure 12 measures how much the bid/ask quotes of TOTF.PA move away from their initial level after a change in the midquote of the future FCE. Since TOTF.PA and FCE are positively correlated, the deviation is positive (resp. negative) for positive (resp. negative) thresholds $\theta$. For $|\theta| \leq 2$, it tends to saturate after approximately 1 second, the same order of magnitude than the lag where the cross-correlation function reaches its maximum. The deviation is quite small, typically less than half a tick, and it increases with $|\theta|$, meaning that the larger the return of the future, the bigger the impact on the stock, which sounds intuitive (see [4] for a similar study on single stock response functions). The curves for $|\theta| > 2$ are a bit messy because of the lack of such events but the qualitative results remain unchanged.

---
[12]Remind that $\delta$ denotes the tick size.

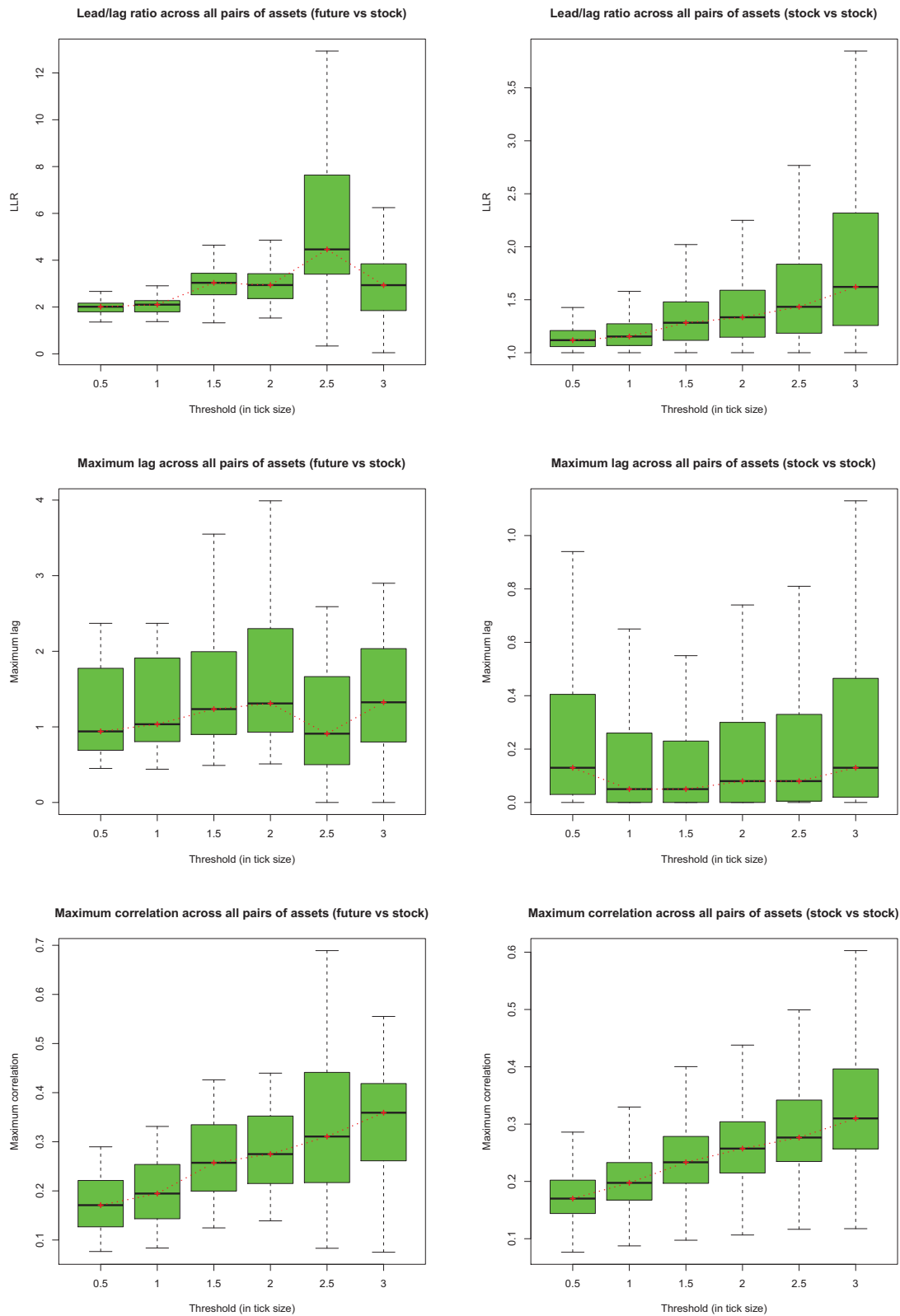

Figure 11: Lead/lag measures as a function of the threshold $\theta$. Left panel: cross-sectional distribution for LLR, maximum lag and maximum correlation for future/stock pairs. Right panel: Idem for stock/stock pairs.

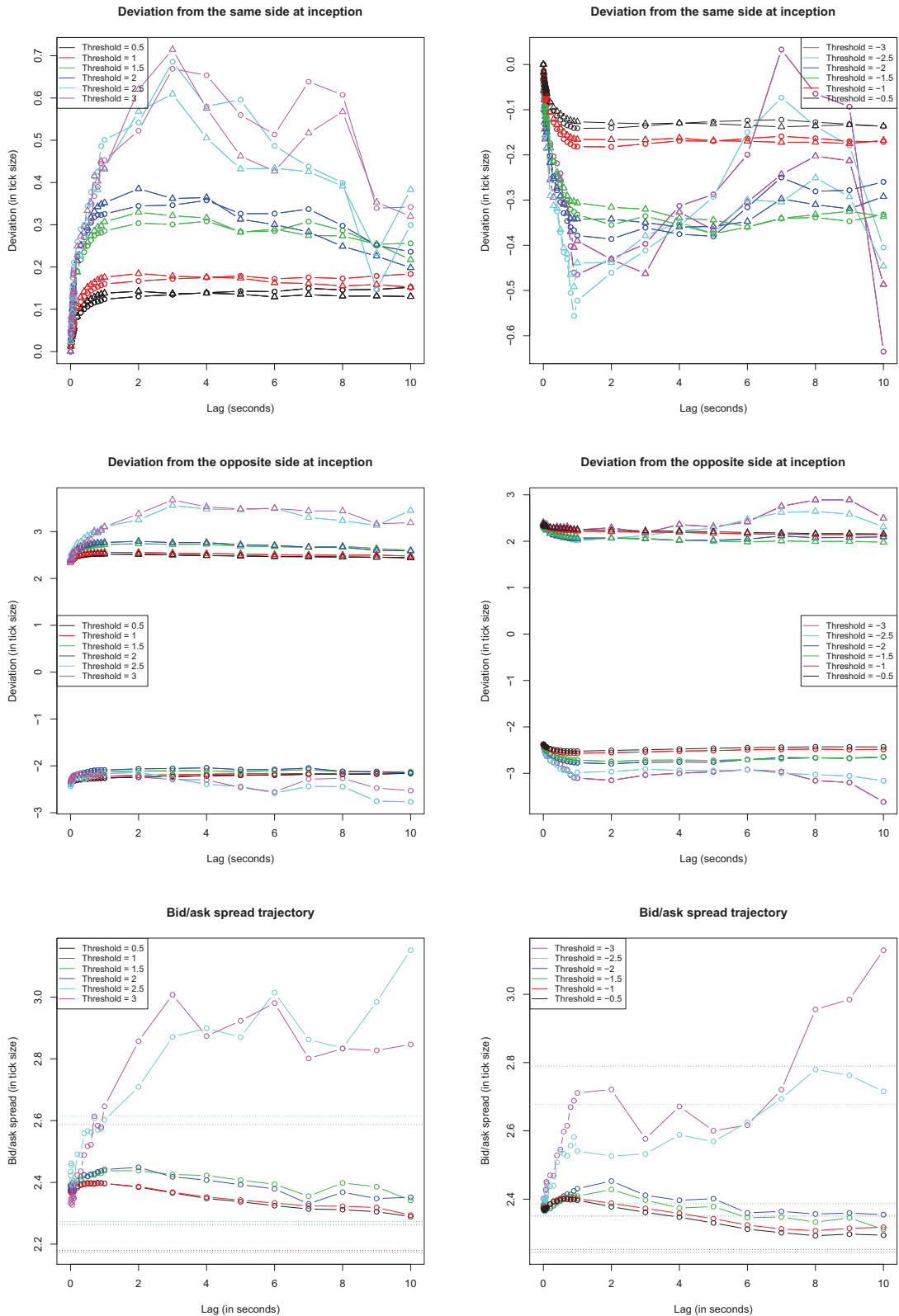

Figure 12: Lead/lag response functions for FCE/TOTF.PA. Top panel: variation of the bid (dots) and ask (triangles) quotes w.r.t. their initial level. Middle panel: variation of the bid (dots) and ask (triangles) quotes w.r.t. the opposite best quote at inception. Bottom panel: variation of the bid/ask spread w.r.t. its initial level. The dotted line is the average bid/ask spread of the stock at time zero.

The middle panel plots the deviation of the bid (resp. ask) quote from the initial best opposite quote, i.e. the ask (resp. bid) quote, after a variation of the midquote of the future. For positive $\theta$ (resp. negative), if the bid (resp. ask) quote becomes larger (resp. smaller) than the initial ask (resp. bid) quote, then it is possible to make money with market orders on the stock by buying (resp. selling) at time zero and unwinding the position afterwards. On the left (resp. right) middle panel, we see that the curve with dots (resp. triangles) is always below (resp. above) zero, so it is not possible to make money with market orders. We would rather lose two ticks on average, which is the average bid/ask spread of TOTF.PA over this period.

Finally, the bottom panel depicts the trajectory of the bid/ask spread after a move of the future. The variation of the spread is not so big for $|\theta| \leq 2$, typically smaller than five percent of the tick size. We observe a relaxation of the spread towards its average value. For $|\theta| > 2$, the spread narrows for small lags before being wider a few seconds after. This can be due to high frequency market making robots of index arbitrage traders, who try to replicate the future with stocks and post quotes accordingly. These people act at very high frequency, often less than a second. The widening of the spread for larger lags might come from agents who follow the evolution of the future as a signal for arbitrage strategies and send market orders on stocks once the future has moved significantly.

### 3.6 Backtest of forecasting devices

The knowledge of lead/lag relationships on financial markets can be used to forecast the short-term evolution of lagging assets and thus to build statistical arbitrage strategies. More precisely, the cross-correlation functions shown in section 3.1 enable us to estimate the direction of the midquote move at the next tick. This forecast is built using the past evolution of the leading asset. For instance, if we assume that $X$ is the leader, and that we are at time $s_{j-1}$, our estimation of the next midquote return of the lagger $r_j^Y = Y_{s_j} - Y_{s_{j-1}}$ is

$$\hat{r}_j^Y = \sum_{k=1}^p \beta_k \sum_i r_i^X \mathbb{1}_{\left\{O_{ij}^{\ell_k} \neq \emptyset\right\}}$$

In the following, we will only be interested in the sign of the midquote return $\text{sign}(\hat{r}_j^Y)$, so we set $\beta_k = \hat{\rho}(\ell_k)$ which is estimated on the last 20 trading days. In practice, we set the last regression lag $\ell_p$ to be the last statistically significant lag. Since our clock is running in tick time, our classification is binary: upward or downward move of the midquote. Note that we also need an estimate of the next tick timestamp of the lagger to compute $\mathbb{1}_{\left\{O_{ij}^{\ell_k} \neq \emptyset\right\}}$. We choose it to be the current timestamp plus the average duration between two ticks over the last 20 trading days.

Once the prediction of the next midquote move is known, if it is positive (resp. negative) we buy (resp. sell) one unit of the lagger, and then we sell (resp. buy) it back after the next tick of the lagger occurs. Regarding the execution costs, we consider two scenarios: execution at the midquote and execution taking into account the bid/ask spread. Midquote execution is clearly not realistic at all but it gives an upper bound for the P&L of the strategy. Note that even with a perfect forecasting device, we need the opposite quote to move more than the initial bid/ask spread at the next tick to make money, which is highly unlikely according to section 3.5.

Figure 13 (resp. 14) plots the accuracy, defined as the percentage of good predictions, of our forecasting device, a random forecast and a forecast based on the autocorrelation function of TOTF.PA (resp. ESSI.PA[13]) over the 44 test days if we take the future FCE to be the leader. It also shows the probabil-ity distribution of returns of the strategy[14]. Figure 15 shows the distribution of returns when taking into account the bid/ask spread[15]. The lead/lag prediction is right in 60.3% (resp. 63.6%) of cases on average for TOTF.PA (resp. ESSI.PA), which is much better than a random forecast[16]. As a result, being able to trade at the mid price yields a profitable strategy with an average return of 0.4 (resp. 0.54) bps per trade. The average daily return is 24.2% (resp. 6.1%) and the standard deviation is 20.2% (resp. 4.6%), thus the annualized Sharpe ratio is 19 (resp. 21). The distribution of returns of the lead/lag strategy is significantly different from the random strategy as judged by the Kolmogorov-Smirnov test, with a distance $D = 0.0936$ (resp. $0.1157$)[17] yielding a p-value of the order of $10^{-16}$. The lead/lag strategy also performs better than the autocorrelation strategy (accuracy of 56.7% for TOTF.PA and 59% for ESSI.PA), itself performing better than a purely random strategy. The Student t-test concludes that both average prediction rates are significantly different from each other (p-value=$2 \times 10^{-11}$ (resp. $3 \times 10^{-7}$)). This shows that using the past information from the leader yields a significant improvement in the forecasting process. This is close to the notion of Granger causality [10]. However, all the profit made by the lead/lag strategy is lost when taking into account the bid/ask spread: the average return is $-2.73$ (resp. $-3.44$) bps per trade.

---

[13]We choose TOTF.PA and ESSI.PA because TOTF.PA is highly liquid in contrary to ESSI.PA.

In order to bypass massive losses due to the bid/ask spread, we can try to predict the midquote of the lagger at a longer horizon than the next tick. For instance, we can sample the data on a bigger tick time basis, i.e. sampling data once the midquote has moved of $\theta$ ticks, with $\theta > 0.5$. Typically, we need to have a tick time bigger than the bid/ask spread, which is 1.88 (resp. 2.83) ticks on average for TOTF.PA (resp. ESSI.PA). We can use the cross-correlation computed at this time scale to forecast the midquote of the lagger. Figure 16 plots the accuracy and the distribution of returns for $\theta = i/2, i = 1, \ldots, 6$. Clearly, the forecasting accuracy deteriorates and the distribution of returns gets wider as the time scale increases, in agreement with the absence of arbitrage. Note that the median return remains negative for any $\theta$.

---

[14]The flame-like shape of the probability distribution comes from the fact that returns can be written as $\frac{m_1-m_0}{m_0} = \frac{i\delta/2}{m_0}$ where $i \in \mathbb{N}$ is the midquote variation expressed in half-ticks and $m_0 \in \mathbb{R}^+$ is the midquote at the inception of the trade.
[15]That is buying at the ask price and selling at the bid price.
[16]The Student t-test for equality of the average prediction rates yields a p-value of $10^{-16}$ for both stocks.
[17]The sample size is 226347 (resp. 241340) returns over the 44 test days.

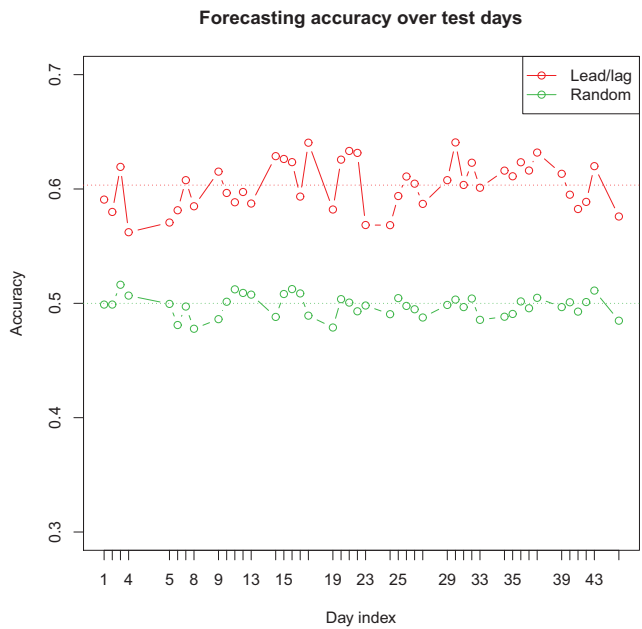
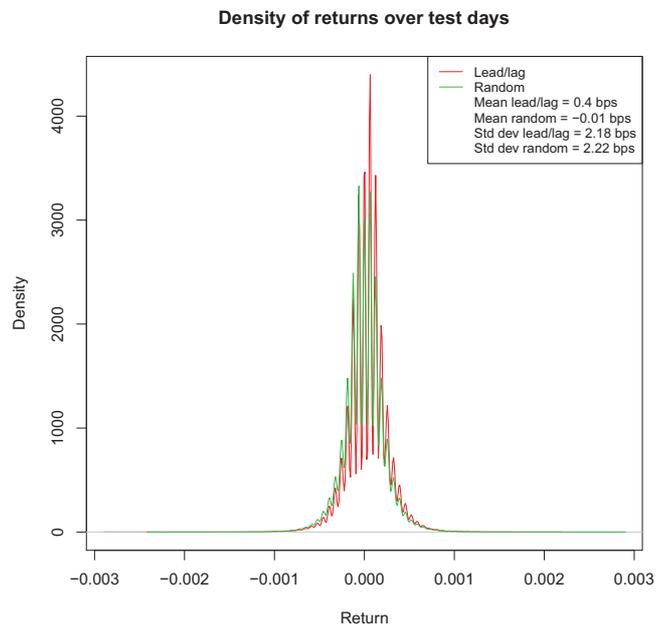
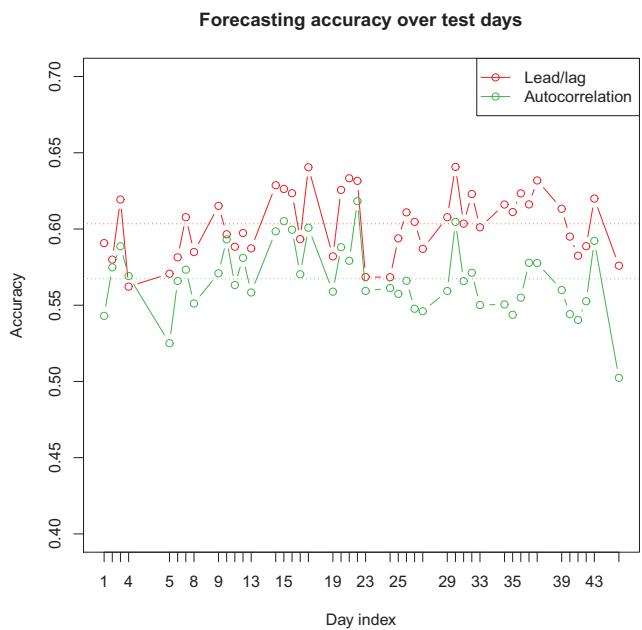
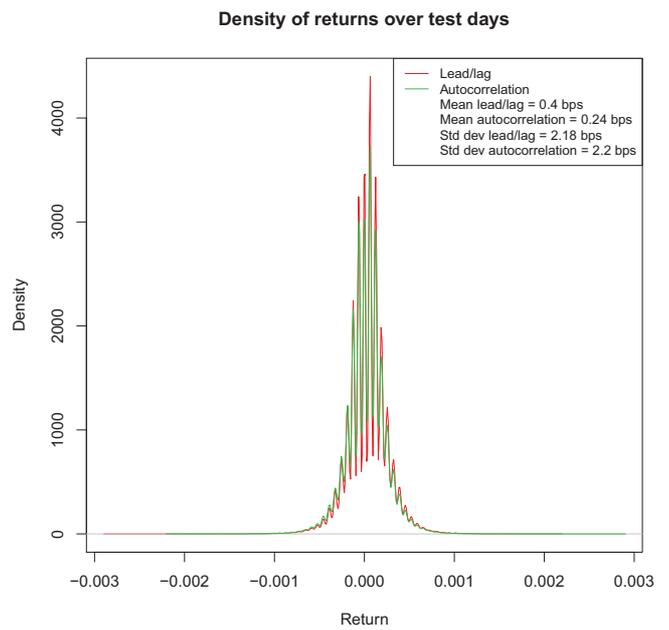

Figure 13: Backtest of the lead/lag strategy (*versus* a random forecast and a forecast based on autocorrelation) on the pair FCE/TOTF.PA over the 44-day test period 2010/03/29 − 2010/05/31. Top left panel: Forecasting accuracy lead/lag *vs* random. Top right panel: Density of the returns of the strategy lead/lag *vs* random. Bottom left panel: Forecasting accuracy lead/lag *vs* autocorrelation. Bottom right panel: Density of the returns of the strategy lead/lag *vs* autocorrelation.

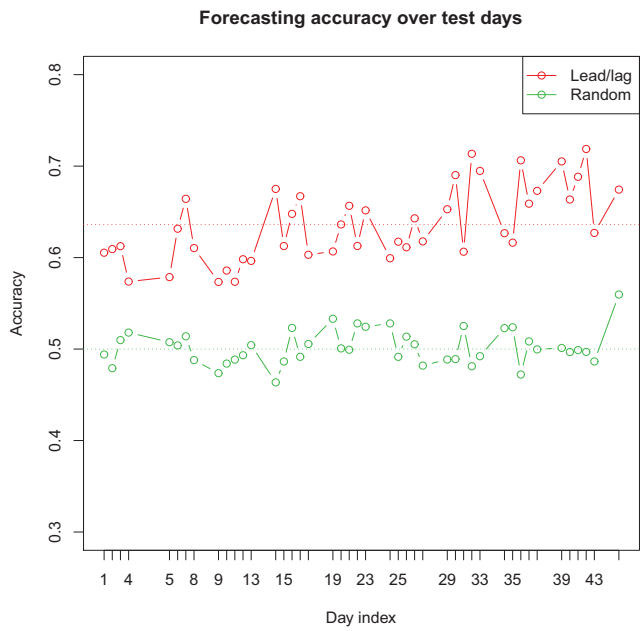
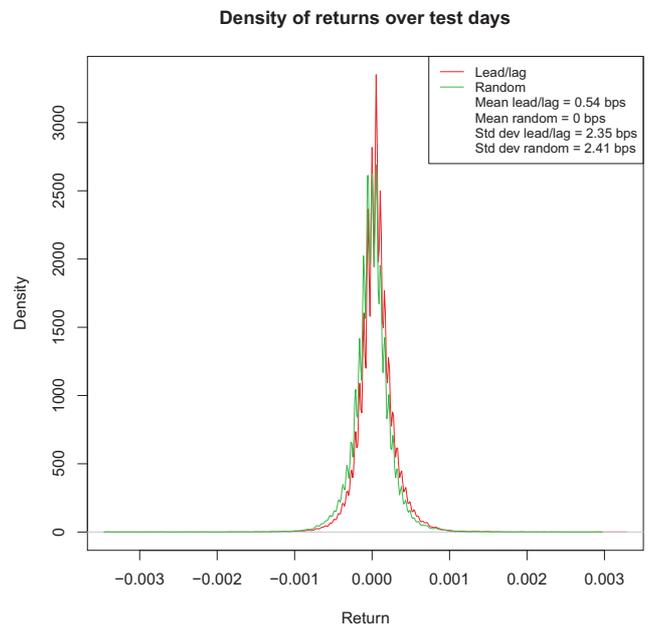
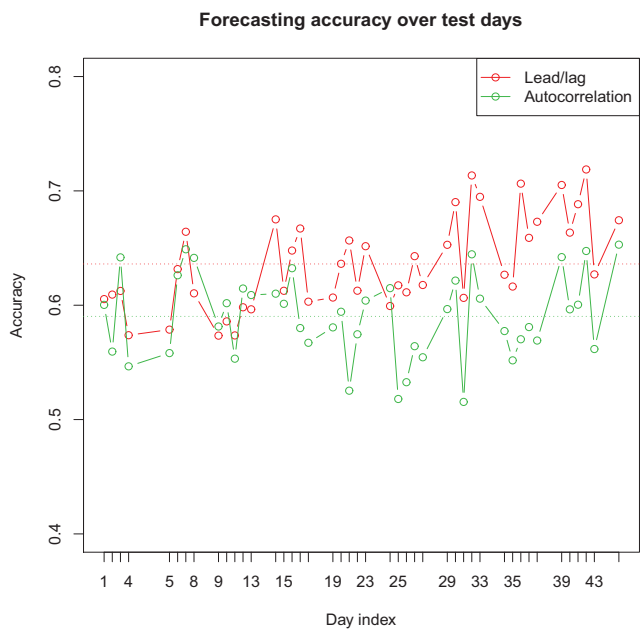
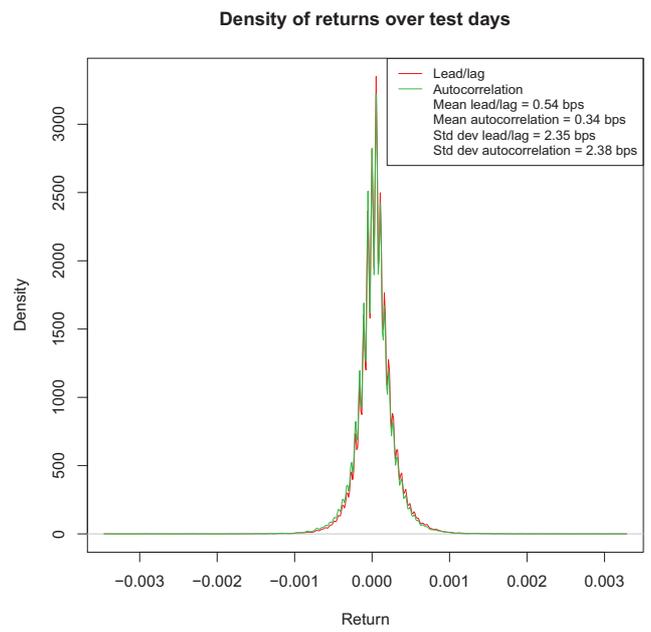

Figure 14: Same as figure 13 for FCE/ESSI.PA.

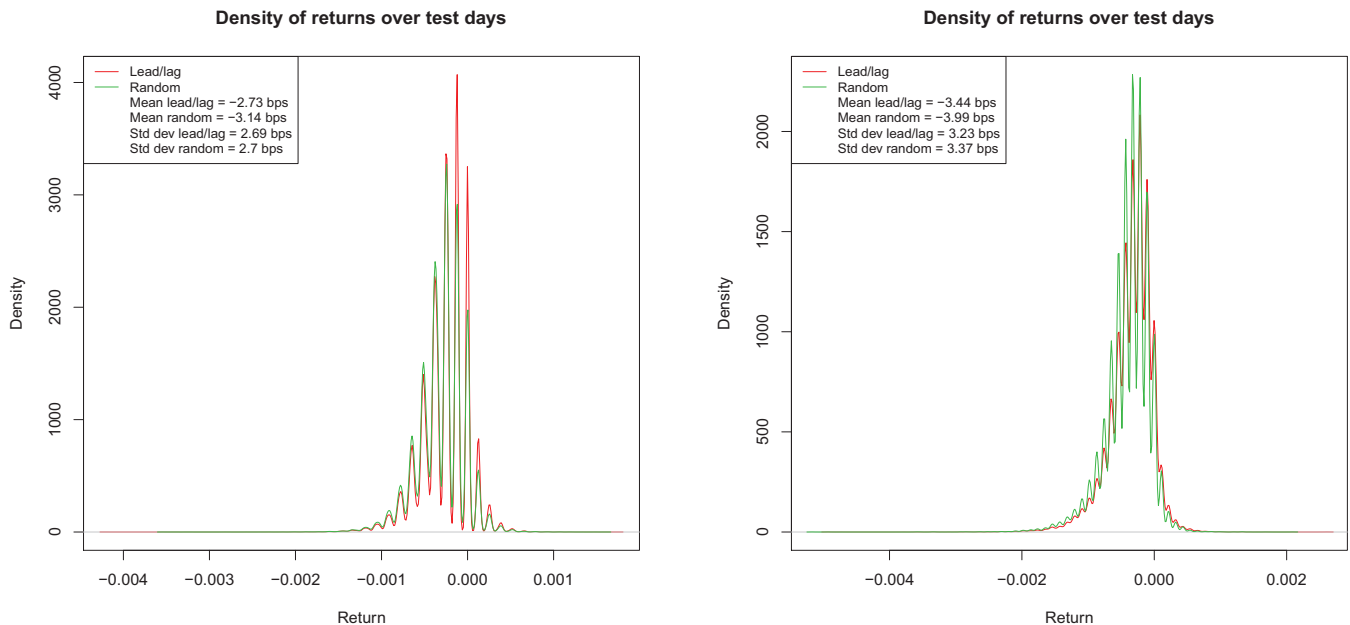

Figure 15: Density of the returns of the lead/lag strategy over the 44-day test period $2010/03/29-2010/05/31$ taking into account the bid/ask spread. Left panel: FCE/TOTF.PA. Right panel: FCE/ESSI.PA

## 4 Conclusion

We study high frequency lead/lag relationships on the French equity market. We use the Hayashi-Yoshida cross-correlation function estimator because it bypasses the issues of asynchrony and artificial liquidity lead/lag. Lead/lag relationships between two stocks or between an equity index future and a stock belonging to this index show different behaviours. The later are far more pronounced than the former. From a more general point of view, we find that the most liquid assets, in terms of short intertrade duration, high trading turnover, narrow bid/ask spread and small volatility tend to lead the others. However, the highest correlations on the market appear for assets displaying similar levels of liquidity. Lead/lag relationships display a non-constant intraday profile, which is different for future/stock and stock/stock pairs. Lead/lag becomes more pronounced, in terms of level of asymmetry and correlation, when focusing on extreme price movements. The study of response functions shows that the average response time of a stock after a move of the index future is of the order of one second and that there is no chance to make money from this effect with market orders. Finally, we obtain an average prediction rate of 60% when forecasting the next midquote variation of a stock with the past evolution of the index future. As said earlier, it does not allow making money by sending market orders only but it could be used for other trading purposes such as market-making or best execution.

In the future, we plan to use the branching structure of Hawkes processes to estimate lead/lag relationships. This is inspired from the declustering algorithm introduced in [22].

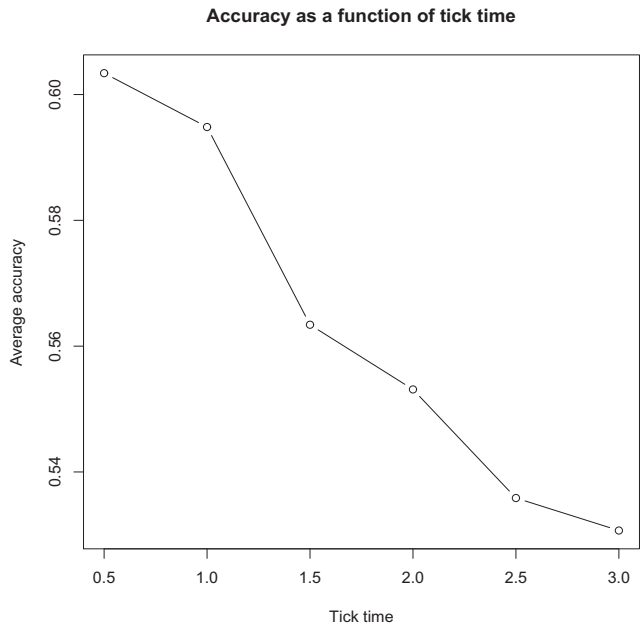
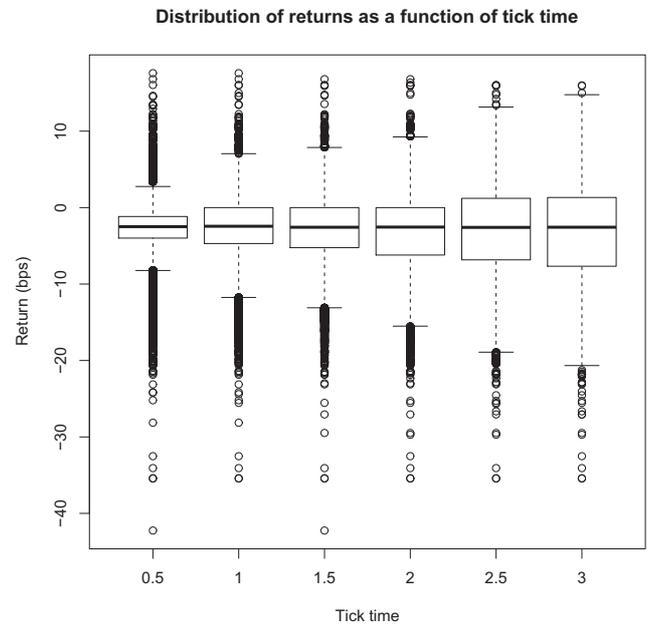
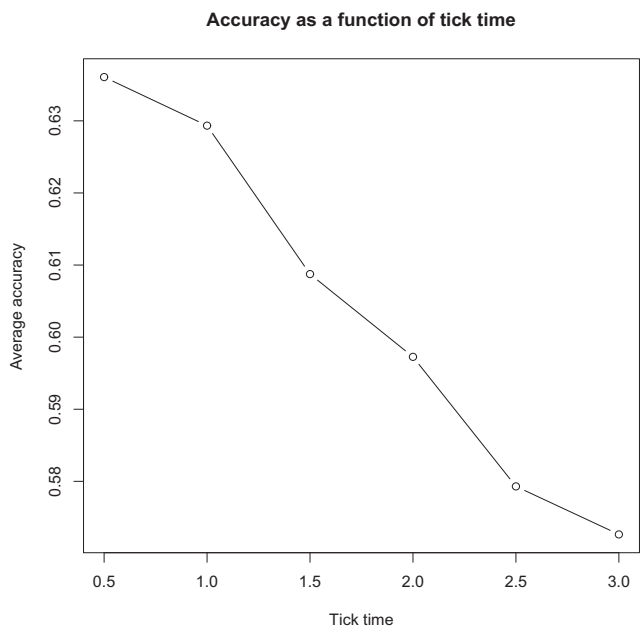
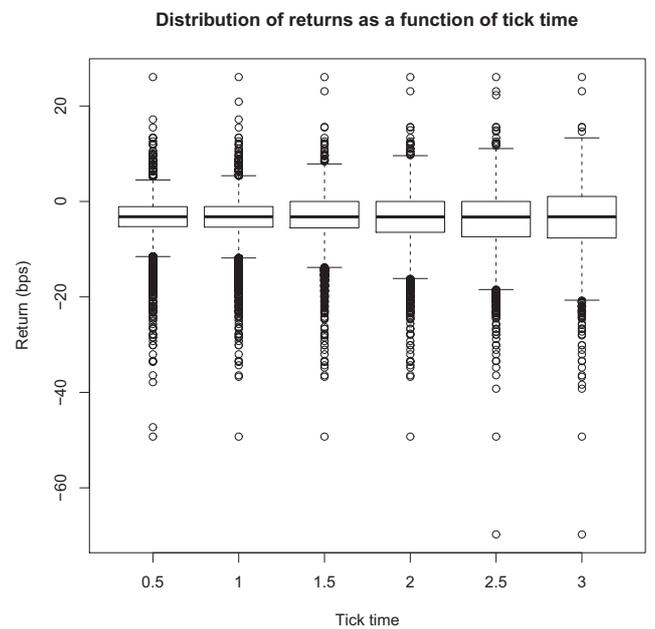

Figure 16: Top left panel: Forecasting accuracy as a function of tick time for FCE/TOTF.PA. Top right panel: Distribution of returns per trade as a function of tick time for FCE/TOTF.PA. Bottom panel: idem for FCE/ESSI.PA

# Appendix A: Explicit computation of LLR

We prove how to end up with the final formulation of the LLR as a ratio of squared correlations. We use the notations introduced in section 2.1. Recall that $X$ leads $Y$ if $X$ forecasts $Y$ more accurately than $Y$ does for $X$,

$$\frac{\|\varepsilon^{YX}\|}{\|r^Y\|} < \frac{\|\varepsilon^{XY}\|}{\|r^X\|}$$

By definition,

$$\varepsilon^{YX} = Y - X\beta = Y - X(C^{XX})^{-1}C^{YX}$$

so that

$$\frac{\|\varepsilon^{YX}\|^2}{\|Y\|^2} = 1 - (C^{YX})^T(C^{XX}C^{YY})^{-1}C^{YX}$$

If we develop $(C^{YX})^T(C^{XX}C^{YY})^{-1}C^{YX}$, we get

$$\frac{\|\varepsilon^{YX}\|^2}{\|Y\|^2} = 1 - \sum_{i,j=1}^{p} C_i^{YX} C_j^{YX} \left(C^{XX}\right)^{-1}_{ij} \left(C^{YY}\right)^{-1}$$

Assuming $C^{XX}$ is diagonal, then we get

$$\frac{\|\varepsilon^{YX}\|^2}{\|Y\|^2} = 1 - \sum_{i=1}^{p} \left(\frac{C_i^{YX}}{\sqrt{C_{ii}^{XX}C^{YY}}}\right)^2$$
$$= 1 - \sum_{i=1}^{p} \rho^2(\ell_i)$$

which ends the computation.

# Appendix B: Explicit computation of $\mathbb{E}(\hat{\rho}(\ell))$

Let us consider two standard Brownian motions $B_1, B_2$ with contemporary correlation $\rho(0) = \rho \in [-1, 1]$. These two Brownian motions are sampled along respective time grids $0 = t_0 \leq t_1 \leq \ldots \leq t_n = T$ and $0 = s_0 \leq s_1 \leq \ldots \leq s_m = T$. The time grids are respectively the jumping times of two independent Poisson processes $N_1$ and $N_2$ and are also independent of $(B_1, B_2)$. This results in piecewise constant processes

$$X(u) = B_1(t(u))$$
$$Y(u) = B_2(s(u))$$
$$t(u) = \max\{t_i | t_i \leq u\}$$
$$s(u) = \max\{s_i | s_i \leq u\}$$

We want to compute $\mathbb{E}(\hat{\rho}(\ell))$ for any lag $\ell$. In fact, we are only interested in the covariance function $\hat{C}(\ell) = \hat{\rho}(\ell)\hat{\sigma}_X\hat{\sigma}_Y$ where $\hat{\sigma}_k^2 = \frac{1}{T}\sum_i (r_i^k)^2$ since standard results[15] show that $\hat{\sigma}_i^2$ is an unbiased and consistent estimator of the realized variance in this framework. Let us assume $\ell \geq 0$. We have, using the notations from section 2.1

$$T.\mathbb{E}(\hat{C}(\ell)) = \mathbb{E}(\sum_{i,j} r_i^X r_j^Y \mathbb{1}_{\{O_{ij}^\ell \neq \emptyset\}})$$

$$= \rho \mathbb{E}(\sum_{i,j} \mathbb{1}_{\{\ell < s_j - t_{i-1}\}} (t_i \wedge s_j - t_{i-1} \vee s_{j-1})^+)$$

where $x^+ = x \vee 0$. Similarly,

$$T.\mathbb{E}(\hat{C}(-\ell)) = \rho \mathbb{E}(\sum_{i,j} \mathbb{1}_{\{\ell < t_i - s_{j-1}\}} (t_i \wedge s_j - t_{i-1} \vee s_{j-1})^+)$$

Let us remark that for $\ell = 0$ the covariance function is unbiased[12] since

$$\mathbb{E}(\sum_{i,j} \mathbb{1}_{\{0 < s_j - t_{i-1}\}} (t_i \wedge s_j - t_{i-1} \vee s_{j-1})^+)$$
$$= \mathbb{E}(\sum_{i,j} (t_i \wedge s_j - t_{i-1} \vee s_{j-1})^+) = \mathbb{E}(T) = T$$

It is also clear that $\mathbb{E}(\hat{C}(T)) = \mathbb{E}(\hat{C}(-T)) = 0$.

$(t_i \wedge s_j - t_{i-1} \vee s_{j-1})^+$ is the length of the overlap between $]t_{i-1}, t_i]$ and $]s_{j-1}, s_j]$. If there is indeed an overlap, it can also be seen as the duration between two consecutive events of the Poisson process resulting from the merge of the two initial Poisson processes. A standard result on the Poisson process[6] states that the merge of two independent Poisson processes is also a Poisson process with an intensity that is the sum of the two. Therefore, we have, for $0 < \ell < T$

$$T.\mathbb{E}(\hat{C}(\ell)) = \rho \mathbb{E}(\sum_{k=1}^{N} \mathbb{1}_{\left\{\ell < \tau_{\bar{i}_2(k)} - \tau_{\underline{i}_1(k)}\right\}} (\tau_k - \tau_{k-1}))$$

where $\{\tau_k, k = 0, \ldots, N = n + m\}$ are the jumping times of the merged Poisson process and

$$\underline{i}_p(k) = \underset{j}{\operatorname{argmax}} \{\tau_j \leq \tau_{k-1} : \tau_j \text{ is of type } p\} \vee 0$$
$$\bar{i}_p(k) = \underset{j}{\operatorname{argmin}} \{\tau_j \geq \tau_k : \tau_j \text{ is of type } p\} \wedge N$$

for $p = 1, 2$. Then, since $N - 1$ is Poisson distributed with parameter $(\lambda_1 + \lambda_2)T$,

$$\mathbb{E}(\hat{C}(\ell)) = \rho e^{-(\lambda_1 + \lambda_2)T} \sum_{n=1}^{+\infty} \frac{((\lambda_1 + \lambda_2)T)^{n-1}}{(n-1)!} \sum_{k=1}^{n} \sum_{i_1=0}^{k-1} \sum_{i_2=k}^{n} \mathbb{P}(\underline{i}_1(k) = i_1 | N = n) \mathbb{P}(\bar{i}_2(k) = i_2 | N = n) f_{k,h}(n, i_1, i_2)$$

$$f_{k,h}(n, i_1, i_2) = \mathbb{E}((\frac{\tau_k}{T} - \frac{\tau_{k-1}}{T}) \mathbb{1}_{\left\{h < \frac{\tau_{i_2}}{T} - \frac{\tau_{i_1}}{T}\right\}} | N = n, \underline{i}_1(k) = i_1, \bar{i}_2(k) = i_2)$$
$$= \frac{(n-1)! \int_{[0,1]^4} (y - x) \mathbb{1}_{\{h < v - u\}} \mathbb{1}_{\{u < x < y < v\}} u^{i_1 - 1} (x - u)^{k - i_1 - 2} (v - y)^{i_2 - 1 - k} (1 - v)^{n - i_2 - 1} \mathrm{d}x \mathrm{d}y \mathrm{d}u \mathrm{d}v}{(i_1 - 1)!(k - i_1 - 2)!(i_2 - 1 - k)!(n - i_2 - 1)!}$$

where $h = \frac{\ell}{T} \in [0, 1[$ and the convention $((-1)!0)^{-1} = 1$ to take into account the boundary cases $i_1 = 0$ and $i_2 = n$. We have used a standard result on the Poisson process[6] which tells that, conditionally to $N_T = n$, the arrival times $t_1, \ldots, t_n$ follow the distribution of the order statistics of the uniform distribution on $[0, T]$. It means that $(u_0 = 0, u_1 = t_1/T, \ldots, u_{n-1} = t_{n-1}/T, u_n = 1)$ has the following probability density function

$$p(u_0, u_1, \ldots, u_{n-1}, u_n) = (n-1)! \mathbb{1}_{\{0 = u_0 < u_1 < \ldots u_{n-1} < u_n = 1\}}$$

which implies that the probability density function of $(u_{i_1}, u_{k-1}, u_k, u_{i_2})$ is

$$p(u_{i_1}, u_{k-1}, u_k, u_{i_2}) = \frac{(n-1)! \mathbb{1}_{\{0 \le u_{i_1} < u_{k-1} < u_k < u_{i_2} \le 1\}} u_{i_1}^{i_1-1} (u_{k-1} - u_{i_1})^{k-i_1-2} (u_{i_2} - u_k)^{i_2-1-k} (1 - u_{i_2})^{n-i_2-1} \delta(u_0) \delta(u_n - 1)}{(i_1 - 1)!(k - i_1 - 2)!(i_2 - 1 - k)!(n - i_2 - 1)!}$$

It is easily seen that

$$\mathbb{P}(\bar{i}_p(k) = i | N = n) = \mathbb{P}(\text{every jump between } \tau_k \text{ and } \tau_{i-1} \text{ is of type } q \text{ and } \tau_i \text{ is a jump of type } p)$$

$$= \frac{\lambda_p}{\lambda_p + \lambda_q} \left(\frac{\lambda_q}{\lambda_p + \lambda_q}\right)^{i-k} \mathbb{1}_{\{k \le i \le n\}} + \delta(i - n) \left(\frac{\lambda_q}{\lambda_p + \lambda_q}\right)^{n-k+1}$$

$$\mathbb{P}(\underline{i}_p(k) = i | N = n) = \mathbb{P}(\text{every jump between } \tau_{i+1} \text{ and } \tau_{k-1} \text{ is of type } q \text{ and } \tau_i \text{ is a jump of type } p)$$

$$= \frac{\lambda_p}{\lambda_p + \lambda_q} \left(\frac{\lambda_q}{\lambda_p + \lambda_q}\right)^{k-i-1} \mathbb{1}_{\{0 \le i \le k-1 < n\}} + \delta(i) \left(\frac{\lambda_q}{\lambda_p + \lambda_q}\right)^k$$

for $p \ne q$. Thus,

$$\mathbb{E}(\hat{C}(\ell)) = \rho e^{-(\lambda_1+\lambda_2)T} \frac{\lambda_1 \lambda_2}{\lambda_1 + \lambda_2} \sum_{n=3}^{+\infty} \frac{((\lambda_1+\lambda_2)T)^{n-1}}{(n-1)!} \sum_{k=2}^{n-1} \sum_{i_1=1}^{k-1} \sum_{i_2=k}^{n-1} \frac{\lambda_1^{i_2-k} \lambda_2^{k-i_1-1}}{(\lambda_1+\lambda_2)^{i_2-i_1}} f_{k,h}(n, i_1, i_2)$$

$$+ \rho e^{-(\lambda_1+\lambda_2)T} \frac{\lambda_2}{\lambda_1 + \lambda_2} \sum_{n=2}^{+\infty} \frac{((\lambda_1+\lambda_2)T)^{n-1}}{(n-1)!} \sum_{k=1}^{n-1} \sum_{i_2=k}^{n-1} \frac{\lambda_1^{i_2-k} \lambda_2^{k-1}}{(\lambda_1+\lambda_2)^{i_2-1}} f_{k,h}(n, 0, i_2)$$

$$+ \rho e^{-(\lambda_1+\lambda_2)T} \frac{\lambda_1}{\lambda_1 + \lambda_2} \sum_{n=2}^{+\infty} \frac{((\lambda_1+\lambda_2)T)^{n-1}}{(n-1)!} \sum_{k=2}^{n} \sum_{i_1=1}^{k-1} \frac{\lambda_1^{n-k} \lambda_2^{k-i_1-1}}{(\lambda_1+\lambda_2)^{n-i_1-1}} f_{k,h}(n, i_1, n)$$

$$+ \rho e^{-(\lambda_1+\lambda_2)T} \sum_{n=1}^{+\infty} \frac{((\lambda_1+\lambda_2)T)^{n-1}}{(n-1)!} \sum_{k=1}^{n} \frac{\lambda_1^{n-k} \lambda_2^{k-1}}{(\lambda_1+\lambda_2)^{n-1}} f_{k,h}(n, 0, n)$$

Similarly,

$$\mathbb{E}(\hat{C}(-\ell)) = \rho e^{-(\lambda_1+\lambda_2)T} \frac{\lambda_1 \lambda_2}{\lambda_1 + \lambda_2} \sum_{n=3}^{+\infty} \frac{((\lambda_1+\lambda_2)T)^{n-1}}{(n-1)!} \sum_{k=2}^{n-1} \sum_{i_1=k}^{n-1} \sum_{i_2=1}^{k-1} \frac{\lambda_1^{k-i_2-1} \lambda_2^{i_1-k}}{(\lambda_1+\lambda_2)^{i_1-i_2}} \tilde{f}_{k,h}(n, i_1, i_2)$$

$$+ \rho e^{-(\lambda_1+\lambda_2)T} \frac{\lambda_1}{\lambda_1 + \lambda_2} \sum_{n=2}^{+\infty} \frac{((\lambda_1+\lambda_2)T)^{n-1}}{(n-1)!} \sum_{k=1}^{n-1} \sum_{i_1=k}^{n-1} \frac{\lambda_1^{k-1} \lambda_2^{i_1-k}}{(\lambda_1+\lambda_2)^{i_1-1}} \tilde{f}_{k,h}(n, i_1, 0)$$

$$+ \rho e^{-(\lambda_1+\lambda_2)T} \frac{\lambda_2}{\lambda_1 + \lambda_2} \sum_{n=2}^{+\infty} \frac{((\lambda_1+\lambda_2)T)^{n-1}}{(n-1)!} \sum_{k=2}^{n} \sum_{i_2=1}^{k-1} \frac{\lambda_1^{k-i_2-1} \lambda_2^{n-k}}{(\lambda_1+\lambda_2)^{n-i_2-1}} \tilde{f}_{k,h}(n, n, i_2)$$

$$+ \rho e^{-(\lambda_1+\lambda_2)T} \sum_{n=1}^{+\infty} \frac{((\lambda_1+\lambda_2)T)^{n-1}}{(n-1)!} \sum_{k=1}^{n} \frac{\lambda_1^{k-1} \lambda_2^{n-k}}{(\lambda_1+\lambda_2)^{n-1}} \tilde{f}_{k,h}(n, n, 0)$$

where

$$\tilde{f}_{k,h}(n,i_1,i_2) = \mathbb{E}((\frac{\tau_k}{T} - \frac{\tau_{k-1}}{T})\mathbb{1}_{\{h < \frac{\tau_{i_1}}{T} - \frac{\tau_{i_2}}{T}\}}|N=n, \bar{i}_1(k)=i_1, \underline{i}_2(k)=i_2)$$

$$= \frac{(n-1)! \int_{[0,1]^4} (y-x)\mathbb{1}_{\{h<v-u\}}\mathbb{1}_{\{u<x<y<v\}} u^{i_2-1}(x-u)^{k-i_2-2}(v-y)^{i_1-1-k}(1-v)^{n-i_1-1} dxdydudv}{(i_2-1)!(k-i_2-2)!(i_1-1-k)!(n-i_1-1)!}$$

$$= f_{k,h}(n,i_2,i_1)$$

which leads

$$\mathbb{E}(\hat{C}(-\ell)) = \rho e^{-(\lambda_1+\lambda_2)T} \frac{\lambda_1 \lambda_2}{\lambda_1+\lambda_2} \sum_{n=3}^{+\infty} \frac{((\lambda_1+\lambda_2)T)^{n-1}}{(n-1)!} \sum_{k=2}^{n-1} \sum_{i_1=1}^{k-1} \sum_{i_2=k}^{n-1} \frac{\lambda_1^{k-i_1-1}\lambda_2^{i_2-k}}{(\lambda_1+\lambda_2)^{i_2-i_1}} f_{k,h}(n,i_1,i_2)$$

$$+ \rho e^{-(\lambda_1+\lambda_2)T} \frac{\lambda_1}{\lambda_1+\lambda_2} \sum_{n=2}^{+\infty} \frac{((\lambda_1+\lambda_2)T)^{n-1}}{(n-1)!} \sum_{k=1}^{n-1} \sum_{i_2=k}^{n-1} \frac{\lambda_1^{k-1}\lambda_2^{i_2-k}}{(\lambda_1+\lambda_2)^{i_2-1}} f_{k,h}(n,0,i_2)$$

$$+ \rho e^{-(\lambda_1+\lambda_2)T} \frac{\lambda_2}{\lambda_1+\lambda_2} \sum_{n=2}^{+\infty} \frac{((\lambda_1+\lambda_2)T)^{n-1}}{(n-1)!} \sum_{k=2}^{n} \sum_{i_1=1}^{k-1} \frac{\lambda_1^{k-i_1-1}\lambda_2^{n-k}}{(\lambda_1+\lambda_2)^{n-i_1-1}} f_{k,h}(n,i_1,n)$$

$$+ \rho e^{-(\lambda_1+\lambda_2)T} \sum_{n=1}^{+\infty} \frac{((\lambda_1+\lambda_2)T)^{n-1}}{(n-1)!} \sum_{k=1}^{n} \frac{\lambda_1^{k-1}\lambda_2^{n-k}}{(\lambda_1+\lambda_2)^{n-1}} f_{k,h}(n,0,n)$$

We now need to compute the integral function $f_{k,h}(n,i_1,i_2)$. After integrating over $y$, we have

$$f_{k,h}(n,i_1,i_2) = \frac{(n-1)! \int_{[0,1]^3} \mathbb{1}_{\{h<v-u\}}\mathbb{1}_{\{u<x<v\}} u^{i_1-1}(x-u)^{k-i_1-2}(v-x)^{i_2-k+1}(1-v)^{n-i_2-1} dxdudv}{(i_1-1)!(k-i_1-2)!(i_2+1-k)!(n-i_2-1)!}$$

$$= \frac{(n-1)! \int_{[0,1]^2} (\int_u^v (x-u)^{k-i_1-2}(v-x)^{i_2-k+1} dx)\mathbb{1}_{\{h<v-u\}}\mathbb{1}_{\{u<v\}} u^{i_1-1}(1-v)^{n-i_2-1} dudv}{(i_1-1)!(k-i_1-2)!(i_2+1-k)!(n-i_2-1)!}$$

We use the following lemma that can be proven by successive integration by parts.

**Lemma 1.** Let $k \in \mathbb{N}^*$, $(i_1,i_2) \in \mathbb{N}^2$ such that $i_1 + 1 < k \leq i_2$. Let $(u,v) \in \mathbb{R}^2$ such that $u \leq v$. Then,

$$\int_u^v (x-u)^{k-i_1-2}(v-x)^{i_2-k+1} dx = \frac{(k-i_1-2)!}{\prod_{p=0}^{k-i_1-2}(i_2-k+2+p)}(v-u)^{i_2-i_1}$$

Using this lemma, we get

$$f_{k,h}(n,i_1,i_2) = \frac{(n-1)! \int_{[0,1]^2} \mathbb{1}_{\{h<v-u\}}\mathbb{1}_{\{u<v\}} u^{i_1-1}(v-u)^{i_2-i_1}(1-v)^{n-i_2-1} dudv}{(i_1-1)!(i_2+1-k)!(n-i_2-1)! \prod_{p=0}^{k-i_1-2}(i_2-k+2+p)}$$

$$= \frac{(n-1)! \int_h^1 (\int_0^{v-h} u^{i_1-1}(v-u)^{i_2-i_1} du)(1-v)^{n-i_2-1} dv}{(i_1-1)!(i_2-i_1)!(n-i_2-1)!}$$

We now use the following lemma, that can also be proven by successive integration by parts.

**Lemma 2.** Let $(i_1,i_2) \in \mathbb{N}^2$ such that $i_1 < i_2$. Let $(h,v) \in \mathbb{R}^2$ such that $0 \leq h < v$. Then,

$$\int_0^{v-h} u^{i_1-1}(v-u)^{i_2-i_1} du = \frac{(i_1-1)!}{\prod_{p=1}^{i_1}(i_2-i_1+p)} v^{i_2} - \sum_{p=1}^{i_1} \frac{\prod_{m=1}^{p-1}(i_1-m)}{\prod_{m=1}^{p}(i_2-i_1+m)} h^{i_2-i_1+p}(v-h)^{i_1-p}$$

We now have

$$f_{k,h}(n,i_1,i_2) = -\frac{(n-1)!}{(i_1-1)!(i_2-i_1)!(n-i_2-1)!}\sum_{p=1}^{i_1}\frac{\prod_{m=1}^{p-1}(i_1-m)}{\prod_{m=1}^{p}(i_2-i_1+m)}h^{i_2-i_1+p}\int_h^1(1-v)^{n-i_2-1}(v-h)^{i_1-p}dv$$

$$+\frac{(n-1)!}{i_2!(n-i_2-1)!}\int_h^1(1-v)^{n-i_2-1}v^{i_2}dv$$

We need to use the two following lemmas. The first one can be proven by the change of variable $u=\frac{v-h}{1-h}$ and the second one by succcessive integration by parts.

**Lemma 3.** *Let $(p,i_1,i_2,n)\in\mathbb{N}^4$ such that $0<p\leq i_1<i_2<n$. Let $h\in\mathbb{R}$ such that $h<1$. Then,*

$$\int_h^1(1-v)^{n-i_2-1}(v-h)^{i_1-p}dv = \frac{(n-i_2-1)!(i_1-p)!}{(n-p-(i_2-i_1))!}(1-h)^{n-p-(i_2-i_1)}$$

**Lemma 4.** *Let $(i_2,n)\in\mathbb{N}^2$ such that $i_2<n$. Let $h\in\mathbb{R}$ such that $h<1$. Then,*

$$\int_h^1(1-v)^{n-i_2-1}v^{i_2}dv = \sum_{p=0}^{i_2}\frac{\prod_{m=1}^{p}(i_2+1-m)}{\prod_{m=0}^{p}(n-i_2+m)}h^{i_2-p}(1-h)^{n-i_2+p}$$

As a result, we have for $h\in[0,1[$

$$f_{k,h}(n,i_1,i_2) = -(n-1)!\sum_{p=1}^{i_1}\frac{h^{i_2-i_1+p}(1-h)^{n-p-(i_2-i_1)}}{(n-p-(i_2-i_1))!(i_2-i_1+p)!}$$

$$+(n-1)!\sum_{p=0}^{i_2}\frac{h^{i_2-p}(1-h)^{n-i_2+p}}{(i_2-p)!(n-i_2+p)!}$$

$$=(n-1)!\sum_{p=i_1}^{i_2}\frac{h^{i_2-p}(1-h)^{n-i_2+p}}{(i_2-p)!(n-i_2+p)!}$$

$$=(n-1)!\sum_{p=0}^{i_2-i_1}\frac{h^{i_2-i_1-p}(1-h)^{n-(i_2-i_1)+p}}{(i_2-i_1-p)!(n-(i_2-i_1)+p)!}$$

$$=g_h(n,i_2-i_1)$$

where $g_h(n,i) = (n-1)!\sum_{p=0}^{i}\frac{h^{i-p}(1-h)^{n-i+p}}{(i-p)!(n-i+p)!} = (n-1)!\sum_{k=0}^{i}\frac{h^k(1-h)^{n-k}}{k!(n-k)!}$. Note that $g_h(n,n) = \frac{1}{n}$ $\forall h$.
Therefore, the average covariance reads

$$\mathbb{E}(\hat{C}(\ell)) = \rho e^{-(\lambda_1+\lambda_2)T}(S_1(\ell)+S_2(\ell)+S_3(\ell)+S_4(\ell))$$

$$S_1(\ell) = \frac{\lambda_1\lambda_2}{\lambda_1+\lambda_2}\sum_{n=3}^{+\infty}\frac{((\lambda_1+\lambda_2)T)^{n-1}}{(n-1)!}\sum_{k=2}^{n-1}\sum_{i_1=1}^{k-1}\sum_{i_2=k}^{n-1}\frac{\lambda_1^{i_2-k}\lambda_2^{k-i_1-1}}{(\lambda_1+\lambda_2)^{i_2-i_1}}g_h(n,i_2-i_1)$$

$$S_2(\ell) = \frac{\lambda_2}{\lambda_1+\lambda_2}\sum_{n=2}^{+\infty}\frac{((\lambda_1+\lambda_2)T)^{n-1}}{(n-1)!}\sum_{k=1}^{n-1}\sum_{i_2=k}^{n-1}\frac{\lambda_1^{i_2-k}\lambda_2^{k-1}}{(\lambda_1+\lambda_2)^{i_2-1}}g_h(n,i_2)$$

$$S_3(\ell) = \frac{\lambda_1}{\lambda_1+\lambda_2}\sum_{n=2}^{+\infty}\frac{((\lambda_1+\lambda_2)T)^{n-1}}{(n-1)!}\sum_{k=2}^{n}\sum_{i_1=1}^{k-1}\frac{\lambda_1^{n-k}\lambda_2^{k-i_1-1}}{(\lambda_1+\lambda_2)^{n-i_1-1}}g_h(n,n-i_1)$$

$$S_4(\ell) = \sum_{n=1}^{+\infty}\frac{((\lambda_1+\lambda_2)T)^{n-1}}{n!}\sum_{k=1}^{n}\frac{\lambda_1^{n-k}\lambda_2^{k-1}}{(\lambda_1+\lambda_2)^{n-1}}$$

Let us consider the case $\lambda_1 \neq \lambda_2$ first. Then,

$$S_1(\ell) = \frac{\lambda_1 \lambda_2}{\lambda_1 + \lambda_2} \sum_{n=3}^{+\infty} \frac{((\lambda_1+\lambda_2)T)^{n-1}}{(n-1)!} \sum_{i_1=1}^{n-2} \sum_{i_2=i_1+1}^{n-1} \frac{\lambda_1^{i_2} \lambda_2^{-i_1-1}}{(\lambda_1+\lambda_2)^{i_2-i_1}} g_h(n, i_2-i_1) \sum_{k=i_1+1}^{i_2} \left(\frac{\lambda_2}{\lambda_1}\right)^k$$
$$= \frac{\lambda_1 \lambda_2}{\lambda_1^2 - \lambda_2^2} \sum_{n=3}^{+\infty} \frac{((\lambda_1+\lambda_2)T)^{n-1}}{(n-1)!} \sum_{i=1}^{n-2} \sum_{j=1}^{n-1-i} \frac{\lambda_1^j - \lambda_2^j}{(\lambda_1+\lambda_2)^j} g_h(n,j)$$

Similarly,

$$S_1(-\ell) = \frac{\lambda_1 \lambda_2}{\lambda_2^2 - \lambda_1^2} \sum_{n=3}^{+\infty} \frac{((\lambda_1+\lambda_2)T)^{n-1}}{(n-1)!} \sum_{i=1}^{n-2} \sum_{j=1}^{n-1-i} \frac{\lambda_2^j - \lambda_1^j}{(\lambda_1+\lambda_2)^j} g_h(n,j)$$
$$= S_1(\ell)$$

and more generally we have $S_i(-\ell) = S_i(\ell)$ for $i \in \{1,2,3,4\}$. Therefore the covariance function is symmetric on average, i.e. $\mathbb{E}(\hat{C}(-\ell)) = \mathbb{E}(\hat{C}(\ell))$. Let us carry on the computations,

$$S_1(\ell) = \frac{\lambda_1 \lambda_2}{\lambda_1^2 - \lambda_2^2} \sum_{n=3}^{+\infty} ((\lambda_1+\lambda_2)T)^{n-1} \sum_{i=1}^{n-2} \sum_{k=1}^{n-1-i} \frac{h^k(1-h)^{n-k}}{k!(n-k)!} \sum_{j=k}^{n-1-i} \frac{\lambda_1^j - \lambda_2^j}{(\lambda_1+\lambda_2)^j} + \frac{(1-h)^n}{n!} \sum_{j=1}^{n-1-i} \frac{\lambda_1^j - \lambda_2^j}{(\lambda_1+\lambda_2)^j}$$

$$= \frac{\lambda_1}{\lambda_1 - \lambda_2} \sum_{n=3}^{+\infty} (\lambda_1+\lambda_2)T)^{n-1} \sum_{k=1}^{n-2} \frac{h^k(1-h)^{n-k}}{k!(n-k)!}(n-1-k)\left(\frac{\lambda_1}{\lambda_1+\lambda_2}\right)^k$$

$$- \frac{\lambda_1}{\lambda_1 - \lambda_2} \sum_{n=3}^{+\infty} (\lambda_1+\lambda_2)T)^{n-1} \left(\frac{\lambda_1}{\lambda_1+\lambda_2}\right)^n \sum_{k=1}^{n-2} \frac{h^k(1-h)^{n-k}}{k!(n-k)!} \sum_{i=1}^{n-1-k} \left(\frac{\lambda_1+\lambda_2}{\lambda_1}\right)^i$$

$$+ \frac{\lambda_1^2}{\lambda_1^2 - \lambda_2^2} \sum_{n=3}^{+\infty} (\lambda_1+\lambda_2)T)^{n-1} \frac{(1-h)^n}{n!}(n-2)$$

$$- \frac{\lambda_1}{\lambda_1 - \lambda_2} \sum_{n=3}^{+\infty} (\lambda_1+\lambda_2)T)^{n-1} \frac{(1-h)^n}{n!}\left(\frac{\lambda_1}{\lambda_1+\lambda_2}\right)^n \sum_{i=1}^{n-2} \left(\frac{\lambda_1+\lambda_2}{\lambda_1}\right)^i$$

$$- \frac{\lambda_2}{\lambda_1 - \lambda_2} \sum_{n=3}^{+\infty} (\lambda_1+\lambda_2)T)^{n-1} \sum_{k=1}^{n-2} \frac{h^k(1-h)^{n-k}}{k!(n-k)!}(n-1-k)\left(\frac{\lambda_2}{\lambda_1+\lambda_2}\right)^k$$

$$+ \frac{\lambda_2}{\lambda_1 - \lambda_2} \sum_{n=3}^{+\infty} (\lambda_1+\lambda_2)T)^{n-1} \left(\frac{\lambda_2}{\lambda_1+\lambda_2}\right)^n \sum_{k=1}^{n-2} \frac{h^k(1-h)^{n-k}}{k!(n-k)!} \sum_{i=1}^{n-1-k} \left(\frac{\lambda_1+\lambda_2}{\lambda_2}\right)^i$$

$$- \frac{\lambda_2^2}{\lambda_1^2 - \lambda_2^2} \sum_{n=3}^{+\infty} (\lambda_1+\lambda_2)T)^{n-1} \frac{(1-h)^n}{n!}(n-2)$$

$$+ \frac{\lambda_2}{\lambda_1 - \lambda_2} \sum_{n=3}^{+\infty} ((\lambda_1+\lambda_2)T)^{n-1} \frac{(1-h)^n}{n!}\left(\frac{\lambda_2}{\lambda_1+\lambda_2}\right)^n \sum_{i=1}^{n-2} \left(\frac{\lambda_1+\lambda_2}{\lambda_2}\right)^i$$

$$S_1(\ell) = \frac{\lambda_1}{\lambda_2(\lambda_1 - \lambda_2)T}(e^{\lambda_1 \ell}(e^{\lambda_1(T-\ell)} - 1 - \lambda_1(T-\ell)) - \frac{1}{2}(\lambda_1(T-\ell))^2)$$
$$- \frac{\lambda_2}{\lambda_1(\lambda_1 - \lambda_2)T}(e^{\lambda_2 \ell}(e^{\lambda_2(T-\ell)} - 1 - \lambda_2(T-\ell)) - \frac{1}{2}(\lambda_2(T-\ell))^2)$$
$$+ \frac{\lambda_1(e^{\lambda_1 \ell} - 1) - \lambda_2(e^{\lambda_2 \ell} - 1)}{(\lambda_1^2 - \lambda_2^2)T}(1 + e^{(\lambda_1+\lambda_2)(T-\ell)}((\lambda_1+\lambda_2)(T-\ell) - 1))$$
$$+ \frac{\frac{\lambda_2^2}{\lambda_1}(e^{\lambda_2 \ell} - 1) - \frac{\lambda_1^2}{\lambda_2}(e^{\lambda_1 \ell} - 1)}{(\lambda_1^2 - \lambda_2^2)T}(e^{(\lambda_1+\lambda_2)(T-\ell)} - 1 - (\lambda_1+\lambda_2)(T-\ell))$$
$$- \frac{\lambda_1^2 + \lambda_2^2}{\lambda_1 \lambda_2 (\lambda_1 + \lambda_2)T}(e^{(\lambda_1+\lambda_2)(T-\ell)} - 1 - (\lambda_1+\lambda_2)(T-\ell) - \frac{((\lambda_1+\lambda_2)(T-\ell))^2}{2})$$
$$+ \frac{(e^{(\lambda_1+\lambda_2)(T-\ell)}((\lambda_1+\lambda_2)(T-\ell) - 2) + (\lambda_1+\lambda_2)(T-\ell) + 2)}{(\lambda_1+\lambda_2)T}$$

Similarly, one can prove that

$$S_2(\ell) = \frac{(e^{(\lambda_1+\lambda_2)(T-\ell)} - 1)}{(\lambda_1 - \lambda_2)T}(e^{\lambda_1 \ell} - 1 - \frac{\lambda_2}{\lambda_1}(e^{\lambda_2 \ell} - 1))$$
$$- \frac{(e^{\lambda_1(T-\ell)} - 1)(e^{\lambda_1 \ell} - 1) + e^{\lambda_1(T-\ell)} - 1 - \lambda_1(T-\ell) - \frac{\lambda_2}{\lambda_1}((e^{\lambda_2(T-\ell)} - 1)(e^{\lambda_2 \ell} - 1) + e^{\lambda_2(T-\ell)} - 1 - \lambda_2(T-\ell))}{(\lambda_1 - \lambda_2)T}$$
$$+ \frac{(e^{(\lambda_1+\lambda_2)(T-\ell)} - 1 - (\lambda_1+\lambda_2)(T-\ell))}{\lambda_1 T}$$
$$S_3(\ell) = \frac{\lambda_1}{\lambda_2} S_2(\ell)$$
$$S_4(\ell) = \frac{e^{\lambda_1 T} - e^{\lambda_2 T}}{(\lambda_1 - \lambda_2)T}$$

The case $\lambda_1 = \lambda_2 = \lambda$ coincides with the limit $\lambda_2 \to \lambda_1$. In this case, we have

$$\mathbb{E}(\hat{C}(\ell)) = \rho e^{-2\lambda T}(S_1(\ell) + 2S_2(\ell) + S_4(\ell))$$
$$S_1(\ell) = e^{\lambda T}(\frac{2}{\lambda T} + 1) - e^{\lambda \ell}(\frac{2}{\lambda T} + \frac{\ell}{T} + (1 - \frac{\ell}{T})(3 + \lambda \ell)) - \frac{2\lambda(T-\ell)^2}{T}$$
$$+ (1 + e^{2\lambda(T-\ell)}(2\lambda(T-\ell) - 1))(\frac{e^{\lambda \ell}(1 + \lambda \ell) - 1}{2\lambda T})$$
$$+ (e^{2\lambda(T-\ell)} - 1 - 2\lambda(T-\ell))(\frac{3 - e^{\lambda \ell}(3 + \lambda \ell)}{2\lambda T})$$
$$- \frac{(e^{2\lambda(T-\ell)} - 1 - 2\lambda(T-\ell) - 2(\lambda(T-\ell))^2)}{\lambda T}$$
$$+ \frac{e^{2\lambda(T-\ell)}(\lambda(T-\ell) - 1) + \lambda(T-\ell) + 1}{\lambda T}$$
$$S_2(\ell) = \frac{(e^{2\lambda(T-\ell)} - 1)}{T}(e^{\lambda \ell}(\ell + \frac{1}{\lambda}) - \frac{1}{\lambda}) + \frac{e^{2\lambda(T-\ell)} - 1 - 2\lambda(T-\ell)}{\lambda T}$$
$$- e^{\lambda T}(1 + \frac{1}{\lambda T}) + \frac{e^{\lambda \ell}}{T}(\frac{1}{\lambda} + \ell) + 2(1 - \frac{\ell}{T})$$
$$S_4(\ell) = e^{\lambda T}$$

## Appendix C: Lead/lag response functions

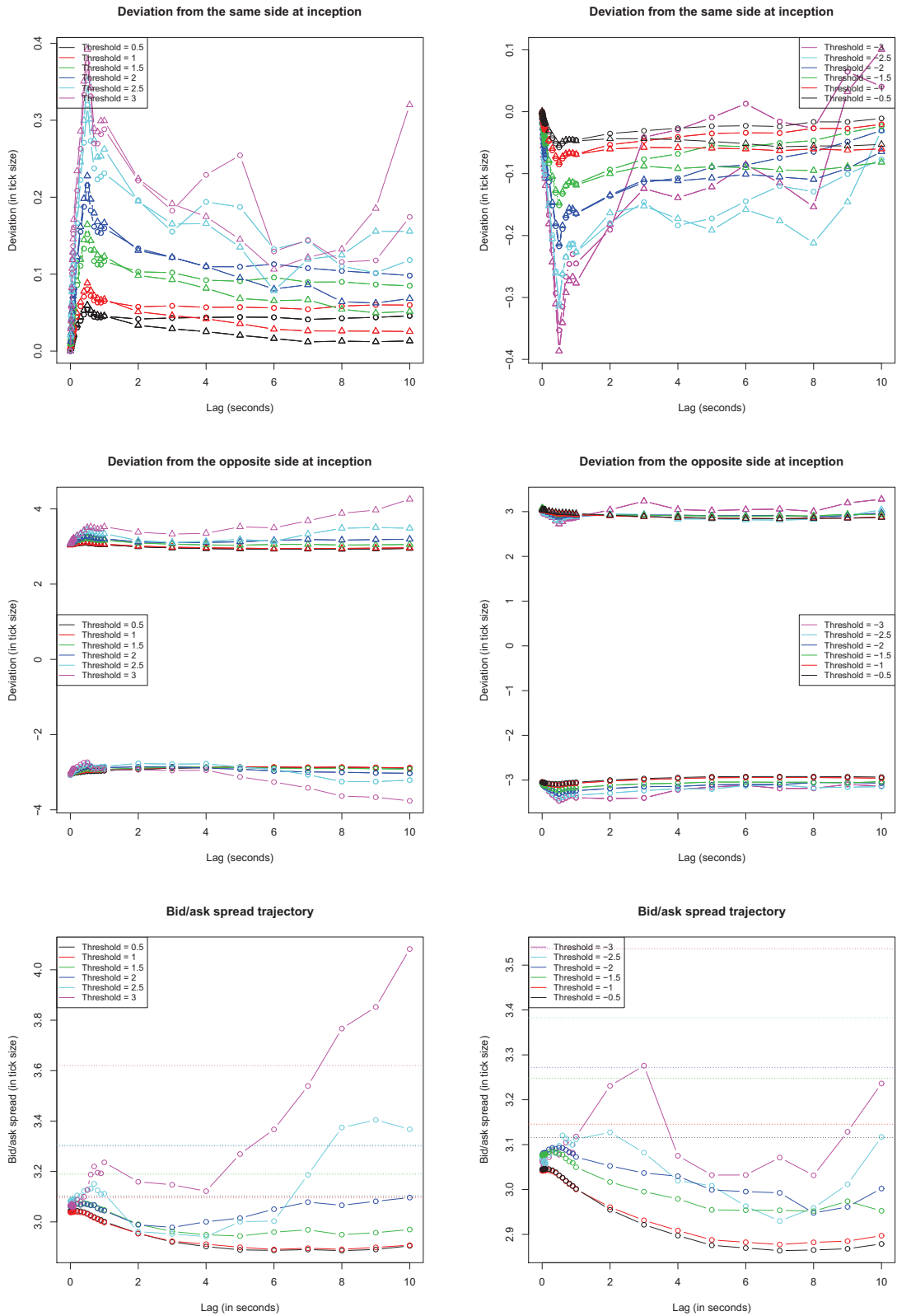

Figure 17: Same as figure 12 for FDX/DTEGn.DE.

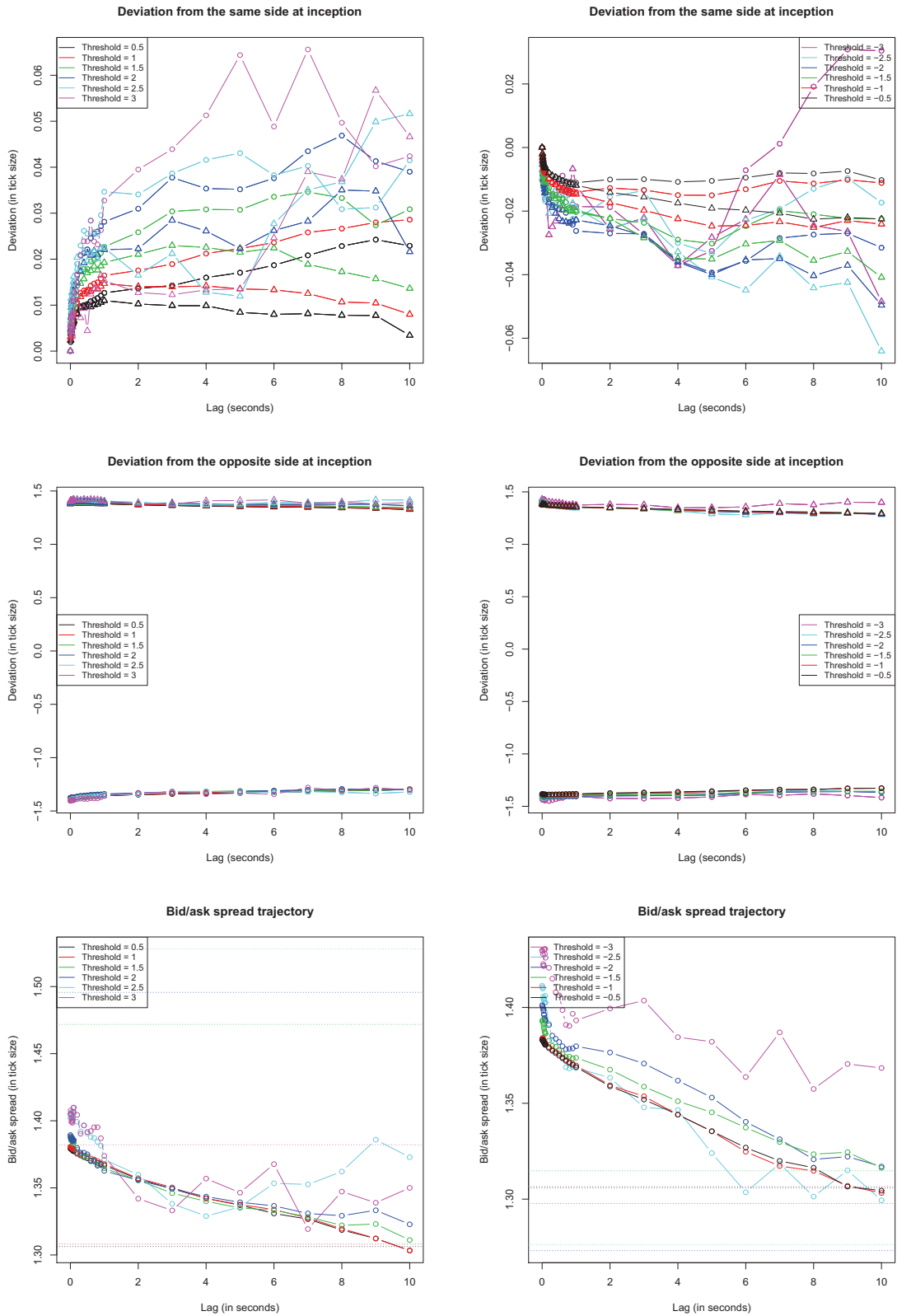

Figure 18: Same as figure 12 for FFI/VOD.L.

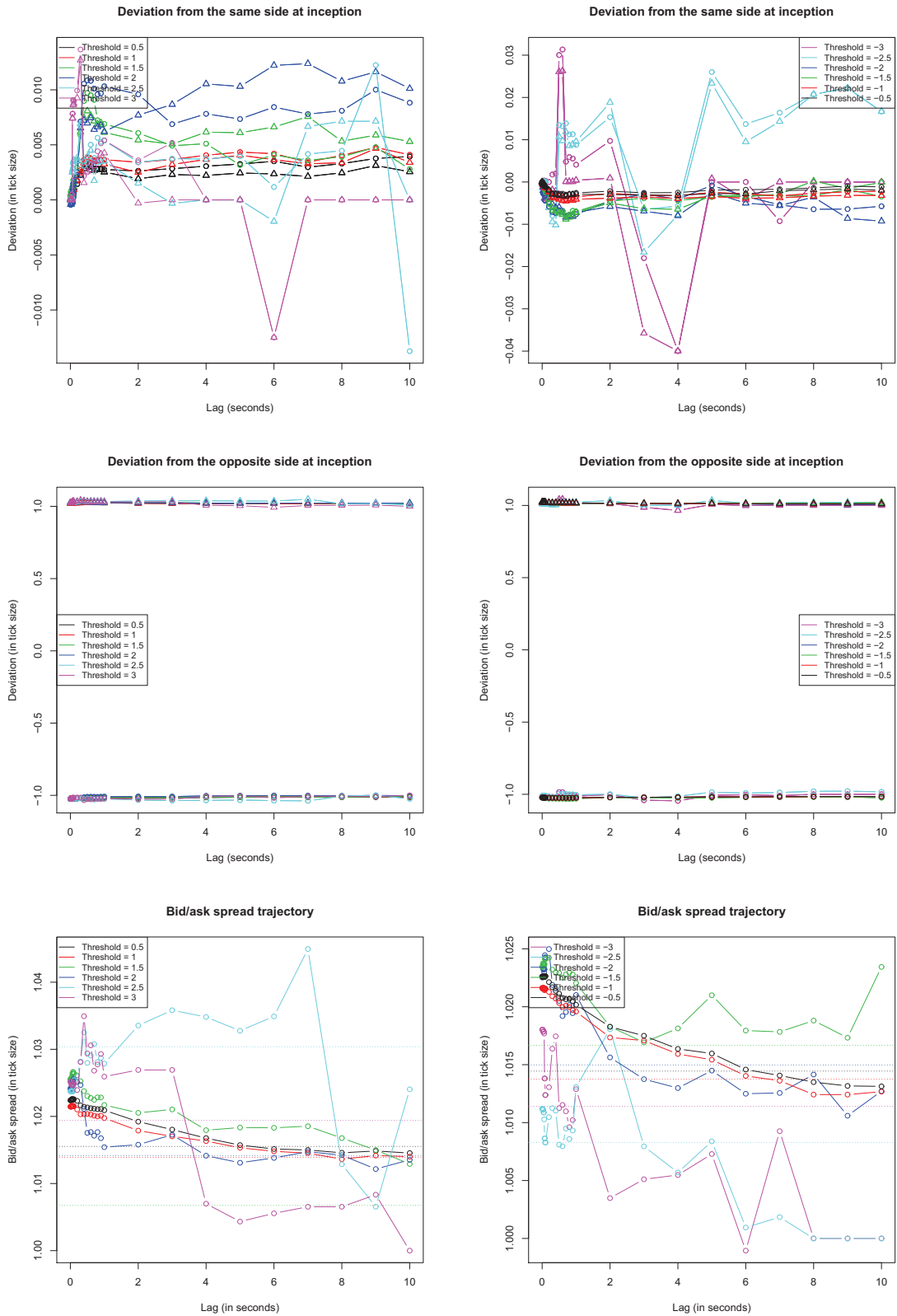

Figure 19: Same as figure 12 for FSMI/NESN.VX.